\documentclass[reprint,amsmath,amssymb,aps,prb]{revtex4-2}

\usepackage[english]{babel}
\usepackage{graphicx}
\usepackage{dcolumn}
\usepackage{bm}
\usepackage{physics}
\usepackage{placeins}
\usepackage{hyperref}
\usepackage[dvipsnames]{xcolor}

\hypersetup{
    colorlinks,
    linkcolor={red!40!black!100!},
    citecolor={blue!30!black!100},
    urlcolor={blue!30!black!100}
}

\begin{document}
\selectlanguage{english}
\preprint{ARK,SR,UniLu}

\title{Understanding electronic excited states in BiFeO$_3$ \textit{via ab initio} calculations and symmetry analysis}

\author{Aseem Rajan Kshirsagar}
\email{aseem.kshirsagar@uni.lu, aseem@posteo.net}

\author{Sven Reichardt}
 
\affiliation{Department of Physics and Materials Science, University of Luxembourg, Luxembourg}

\date{\today}

\begin{abstract}

BiFeO$_3$ is a technologically relevant multiferroic perovskite featuring ferroelectricity and antiferromagnetism. 
Its lattice, magnetic, and ferroelectic degrees of freedoms are coupled to its optically active excitations and thus hold the potential to be reversible probed and controlled by light.
In this work, we combine \textit{ab initio} density functional and many-body perturbation theory methods with an extensive symmetry and atomic-orbital analysis to describe and understand the electronic excited states spectrum and its imprint on the optical absorption spectrum with quantitative accuracy and qualitative insights.
We find that the optical absorption spectrum of BiFeO$_3$ contain several strongly bound and spatially localized electronic transitions in which the spin-degree of freedom is almost fully flipped.
With our analysis we thoroughly characterize these localized spin-flip transitions in terms of the unusual crystal field splitting of Fe-$3d$ single-electron orbitals.
Our symmetry analysis further allows us to thoroughly explain how the spin content and the energetic fine structure of these strongly bound excitons are dictated by the interplay between crystal symmetry, electron-hole attraction, and the spin-orbit coupling. 

\end{abstract}

\maketitle

 \section{Introduction}

Multiferroics, materials which exhibit multiple ferroic orders, such as ferromagnetism and ferroelectricity~\cite{Fiebig2016-sx}, are promising candidates for a variety of technological applications in which the coupling between two different ferroic properties can be exploited to manipulate one by tuning the other.
Among the multiferroics, BiFeO$_3$ stands out as one of the few single-phase, room-temperature multiferroic materials~\cite{Kuo2016}.
Magnetically, BiFeO$_3$ is approximately an antiferromagnet ~\cite{AFM_neel_1,AFM_neel_2,Ederer2005-nf,Meyer2022-ab}.
The antiferromagnetically aligned Fe spins are in addition slightly canted due to an anti-symmetric exchange (Dzyaloshinskii–Moriya) interaction, giving rise to a spin cycloid of period $\sim 62$~nm~\cite{Sosnowska1982,Ratcliff2016-nf,Xu2021-wu}.
BiFeO$_3$ also features ferroelectricity, which manifests itself as a large electric polarization of $\sim 90\mu$C/cm$^2$~\cite{Moreau1971,Catalan2009,wang_epitaxial_2003} along the [111] crystal direction.
The ferroelectric polarization originates from the repulsive interaction between the lone electron pair in the $6s$ shell of Bi and the Fe atoms~\cite{Spaldin2017,Seshadri2001,Neaton2005}, which results in a structural distortion and intrinsic dipole moments. \par 

While BiFeO$_3$ is most celebrated for the demonstrated electric-field control of magnetic domains and spin waves ~\cite{Heron2014-qa,Rovillain2010-kb}, its optical properties and their possible applications have also been explored along various avenues.
The optical excitations in BiFeO$_3$ are subject to the rich interplay of the electronic, lattice, and ferroic degrees of freedom and give rise to phenomena such as light-induced strain~\cite{Kundys2010-fl,Chen_2012,PhysRevLett.110.037601,PhysRevB.85.092301,PhysRevB.102.220303,Li2015,Lee2022}, light-induced switching of ferroic order parameters ~\cite{Yang2018-tx,Liou2019-to}, a bulk photovoltaic effect~\cite{Choi_2009,Yang_2010,PhysRevLett.109.236601,Bhatnagar_2013,You_2018,Knoche_2021}, and photocatalysis~\cite{Li_2010,Bharathkumar_2016,bai_size_2016,Amdouni_2023}.
Lastly, the electronic excitations in BiFeO$_3$ have also been shown to interact with its magnetic texture and allow the generation of spin waves and magnons~\cite{Ramirez2009-rx,ramirez_magnon_2009,xu_optical_2009,Cazayous2009-go}. \par

BiFeO$_3$ is a wide-band gap semiconductor with an optical band gap of 2.7 eV~\cite{sando_revisiting_2018}.
However, optical spectroscopy on BiFeO$_3$ reveals a finite absorption coefficient also for photon energies below 2.7 eV.
Most prominently, the absorption spectrum of BiFeO$_3$ below the band gap features resonance peaks at 1.4 eV and 1.9 eV that have been hypothezised to stem from electronic transitions within the $3d$ shell of the Fe atoms in which the electron spin is flipped~\cite{Galuza1998,xu_optical_2009,gomez-salces_effect_2012,Wei2017-bd,meggle_optical_pressure_2019,ramirez_magnon_2009}.
Additional absorption features have also been observed at photon energies of 2.3 eV and 2.5 eV as a prominent shoulder of the bulk absorption onset, the origin of which has not been clarified yet~\cite{meggle_temperature_2019,meggle_optical_pressure_2019,chen_optical_2010,xu_optical_2009,ramirez_magnon_2009,hauser_characterization_2008,chen_optical_2010,schmidt_anisotropic_2015}, with possible explanations ranging from Fe $3d$ shell transitions~\cite{ramirez_magnon_2009,Ramirez2009-rx}, to defect states~\cite{radmilovic_combined_2020,clark_energy_2009,ju_first-principles_2009,hauser_characterization_2008,moubah_photoluminescence_2012,Jiang2011-tb} to self-trapped excitons~\cite{pisarev_charge_2009}, and even bi-excitons~\cite{ramirez_magnon_2009,schmidt_anisotropic_2015}.
The in-gap optical excitations have also been linked with light-induced optical features~\cite{meggle_temperature_2019,meggle_optical_pressure_2019,Burkert2016}, light induced strain~\cite{Kundys2010-fl,PhysRevB.85.092301}, light-induced magnetic excitations~\cite{xu_optical_2009,ramirez_magnon_2009,Ramirez2009-rx}, and Raman resonance~\cite{Cazayous2009-go,weber_temperature_2016,Ramirez2009-rx,Singh2008-va}.
These excitations induce an interplay of charge, spin, and vibrational degrees of freedom in BiFeO$_3$, and potentially can be used to control these degrees of freedom. Thus, a consistent and clear theoretical description of these excitons is warranted. \par

On the theoretical side, a coherent picture of these electronic excitations below the band gap is still missing.
While the ground state of BiFeO$_3$ has previously been well-studied and -understood with the help of first-principles methods such as density functional theory (DFT)~\cite{Ederer2005-fq,Ravindran2006-eu,Neaton2005,wang_epitaxial_2003,Hermet2007-de}, a successful description of its excited states has proven to be far more challenging.
The use of standard local and semi-local approximations for the electron-electron interaction, such as the local density approximation (LDA)~\cite{lda} and the generalized gradient approximation (GGA)~\cite{Perdew1996}, leads to an excessive delocalization of the Fe-$3d$ electrons, and a consequent poor description of single-electron excited states and their relative energy levels.
Several post-DFT methods, such as the inclusion of a localized, repulsive Hubbard-$U$ potential (DFT+U)~\cite{lda_u} or using so-called hybrid functionals~\cite{hybrid}, which include a fraction of the unscreened exchange interaction, have been applied to BiFeO$_3$ in the past to address these shortcomings.
While these methods do lead to a better prediction for the electronic band gap~\cite{Rahimi2022-ie,Goffinet2009-qk,Ju2009-xb}, they come at the cost of a tunable parameter.
In addition, LDA+U predicts an indirect band gap, compared to a direct band gap in local/semi-local DFT, due to a spurious realignment of the conduction bands involving the Fe-$3d$ orbitals~\cite{shenton_effects_2017,Neaton2005,gillet_ab_2017,Ghosal2022-tc} and is thus not entirely satisfying on its own.
Calculations of the optical absorption spectrum using these post-DFT methods have also been limited to using the independent-particle (IP) approximation for the electronic excitation spectrum~\cite{Rahimi2022-ie,Sando2016-vh,Lima2020-bu,Young2012-te,Yaakob2015-dn,Naz2022-so,Ghosal2022-tc}. 
While this is sufficient to yield a reasonable prediction of the optical absorption spectrum compared to experiment beyond the absorption onset, this approach inherently fails to capture modifications of the spectrum due to the interaction of electron-hole pairs and to capture bound, in-gap states, \textit{i.e.}, excitons. 
In order to better understand the nature of the latter, a theoretical study is needed that takes into account the electron-hole interaction in a more quantitatively predictive way.   \par 

With this work, we aim to present a detailed analysis of the electronic excited states of BiFeO$_3$ and its optical absorption spectrum using state-of-the-art many-body perturbation theory (MBPT) and group theory methods.
Using the \textit{GW} approximation originally proposed by Hedin~\cite{Hedin1965-sv,Hybertsen1985-ne} to correct DFT quasi-particle energies and the Bethe-Salpeter equation (BSE)~\cite{Salpeter1951-va,Rohlfing2000-sm} to account for the interaction between electron-hole pairs and the formation of excitons, we calculate the absorption spectrum of BiFeO$_3$ and analyze it in terms of electronic transitions involved.
We further take a closer look at the crystal field splitting of Fe-$3d$ orbitals and its effect on the energy ordering of excitons.
We scrutinize and analyze the spectrum of excited states in terms of symmetry, spin content, and their overlap with atomic orbitals.
Our analysis elucidates the composition of excitons in intuitive chemical bonding terms and dissects the excitonic fine induced by the spin-orbit interaction.
While our results are in quantitative agreement with the available experimental data, we believe that our work also sheds light on the qualitative nature of strongly bound excited states within the band gap. 
Finally, our findings also suggest that the reported photo-induced in-gap optical features~\cite{meggle_temperature_2019,meggle_optical_pressure_2019,Burkert2016} are linked with Fe-3$d$ transitions. \par 

This paper is accompanied by our companion paper~\cite{aseem_sven_prl}, in which we focus on the spin composition and spatial localization of excitons in BiFeO$_3$ and demonstrate that circularly polarized light can be used to excite chiral excitons to drive a net spin magnetization. 

\subsection{Methods}

We use DFT in the formulation of Kohn and Sham~\cite{ks-dft} to obtain the ground-state electronic properties that serve as the starting point for our subsequent MBPT calculations.
The exchange-correlation potential is approximated at the level of the GGA in the parametrization of Perdew, Burke, and Ernzerhof (PBE)~\cite{Perdew1996}.

We improve the DFT-level quasi-particle band structure by calculating the diagonal elements of the dynamic electron self-energy within the \textit{GW} approximation~\cite{Hedin1965-sv,Onida2002-qg}:
\begin{equation}\label{eq:SigmaGW}
\begin{split}
    \Sigma_{\mathbf{k},n}(\omega) &= \int \mathrm{d}^3 r \int \mathrm{d}^3 r' \int \frac{\mathrm{d} \omega'}{2 \pi} \, \psi^*_{\mathbf{k},n}(\mathbf{r}) \psi_{\mathbf{k},n}(\mathbf{r'}) \\
    & \times i G(\mathbf{r},\mathbf{r}';\omega+\omega') W(\mathbf{r},\mathbf{r}';\omega') 
    - v^{(\mathrm{xc})}_{\mathbf{k},n}.
\end{split}
\end{equation} 
Here, $\psi_{\mathbf{k},n}(\mathbf{r})$ denotes the wave function of the non-interacting Kohn-Sham state of band $n$ and wave vector $\mathbf{k}$, $v^{\mathrm{(xc)}}_{\mathbf{k},n}$ is the corresponding PBE exchange-correlation energy, $W(\mathbf{r},\mathbf{r}';\omega)$ is the dynamically screened Coulomb potential, and $G(\mathbf{r},\mathbf{r'};\omega)$ denotes the one-electron Green's function.
We use the random phase approximation (RPA) to obtain the dynamically screened Coulomb interaction $W$ and approximate its pole structure by a single pole using the Godby-Needs plasmon-pole model~\cite{Oschlies1995} to find a reasonable compromise between accuracy and computational cost.
We construct the one-electron Green's function $G$ needed for the calculation of the self-energy $\Sigma$ and for the irreducible polarizability needed for $W$ from the Kohn-Sham wave functions $\psi_{\mathbf{k},n}(\mathbf{r})$ and quasi-particle energies $E_{\mathbf{k},n}$.
We obtain the latter in a self-consistent way from the relation
\begin{equation}\label{eq:evGW}
    E_{\mathbf{k},n} = \varepsilon^{(\mathrm{KS})}_{\mathbf{k},n}
    + \Sigma_{\mathbf{k},n}(E_{\mathbf{k},n})[G(\{E_{\mathbf{k},n}\}),W(\{E_{\mathbf{k},n}\})],
\end{equation}
starting from the Kohn-Sham quasi-particle energies $\varepsilon^{(\mathrm{KS})}_{\mathbf{k},n}$ as an initial guess for $E_{\mathbf{k},n}$.
This eigenvalue self-consistent approximation of the $GW$ self-energy (``ev$GW$'') has been shown to be more accurate than the single-iteration version obtained by setting $E_{\mathbf{k},n}=\varepsilon^{(\mathrm{KS})}_{\mathbf{k},n}$ (``$G_0 W_0$'')~\cite{Shishkin2007-hj}.

To obtain the electronic excitation and absorption spectrum beyond the IP-level, we construct and solve the BSE in the basis of transitions between occupied and empty Kohn-Sham states using the Kohn-Sham wave functions and the ev$GW$ eigenvalues~\cite{Salpeter1951-va,Strinati1982-zz,Rohlfing2000-sm,Palummo2004-xk}.
We approximate the electron-hole interaction kernel as the sum of an attractive Coulomb potential screened by the static dielectric function in the RPA and the repulsive, unscreened exchange interaction.
We work within the Tamm-Dancoff  approximation~\cite{Tamm1991-cv,Dancoff1950-qo} and neglect the coupling of positive and negative energy transitions.
The solution of the BSE yields the exciton energies $E_S$ and exciton envelop wave functions $A^S_{\mathbf{k},c,v} = \langle \mathbf{k},v \to \mathbf{k},c | S \rangle$, which represent the overlap of the exciton state $|S\rangle$ with the IP-transition from an occupied, valence band state $|\mathbf{k},v\rangle$ to an empty, conduction band state $|\mathbf{k},c\rangle$.

From the exciton states and energies, we calculate the macroscopic dielectric function along direction Cartesian direction $i$ via ~\cite{Martin2016-fb,Rohlfing2000-sm} 
\begin{equation}\label{eq:eps_bse}
    \epsilon_i(\omega) = 1 - \frac{4\pi e^2}{V_{\mathrm{uc}}} \sum_S
    |\langle S|\hat{r}_i|0\rangle|^2 \frac{2(E_S - i\gamma)}
    {(\hbar\omega)^2 - (E_S - i\gamma)^2},
\end{equation}
where $e$ denotes the magnitude of the electron charge in Gaussian units and $V_{\mathrm{uc}}$ is the volume of the unit cell of BiFeO$_3$.
The matrix elements of the $i$th component of the position operator between the ground state and the exciton states are given by $\langle S|\hat{r}_i|0\rangle = \sum_{\mathbf{k},c,v} \left(A^S_{\mathbf{k},c,v}\right)^* \langle\mathbf{k},c|\hat{r}_i|\mathbf{k},v\rangle$ where the last factor is the matrix element of the position operator between a valence band state $|\mathbf{k},v\rangle$ and a conduction band state $|\mathbf{k},c\rangle$.
For simplicity, we use a constant electronic broadening of $\gamma=100$~meV for all exciton states.
We note that the summation over bands implicitly contains a summation over the spin degree of freedom, \textit{i.e.}, the summation over exciton states $S$ in Eq.~\eqref{eq:eps_bse} would yield a factor of 2 in case of spin-degenerate bands.
However, since spin-orbit coupling leads to a sizable mixing of states of different spin polarization, the summation over the spin degree of freedom does not simply yield a factor of 2 here.

Lastly, we compare our state-of-the-art $GW$+BSE calculations to the predictions obtained on the IP-level using both the PBE wave functions and quasi-particle energies, as well as those from PBE+U, \textit{i.e.}, including a Hubbard-$U$ interaction correction~\cite{lda_u,Anisimov1997-sj}.
In this latter case, as the additional $U$-term already leads to an optical gap quantitatively comparable to experiment, we calculate the macroscopic dielectric function on the IP-level:
\begin{equation}\label{eq:eps_ip}
\begin{split}
    \epsilon_i(\omega) = 1 - \frac{4\pi e^2}{V_{\mathrm{uc}}} \sum_{\mathbf{k},c,v} |\langle \mathbf{k},c|\hat{r}_i|  \mathbf{k},v\rangle|^2  \\ \times \frac{2(\Delta\varepsilon_{\mathbf{k},c,v} - i\gamma)} {(\hbar\omega)^2 - (\Delta\varepsilon_{\mathbf{k},c,v} - i\gamma)^2}
\end{split}
\end{equation}
in which $\Delta\varepsilon_{\mathbf{k},c,v}=\varepsilon^{(\mathrm{KS})}_{\mathbf{k},c}-\varepsilon^{(\mathrm{KS})}_{\mathbf{k},v}$ is the difference of the Kohn-Sham quasi-particle energies of valence and conduction band states at the same wave vector.
 
\subsection{Computational details}

\begin{figure}[]
    \centering
    \includegraphics[width=0.48\textwidth]{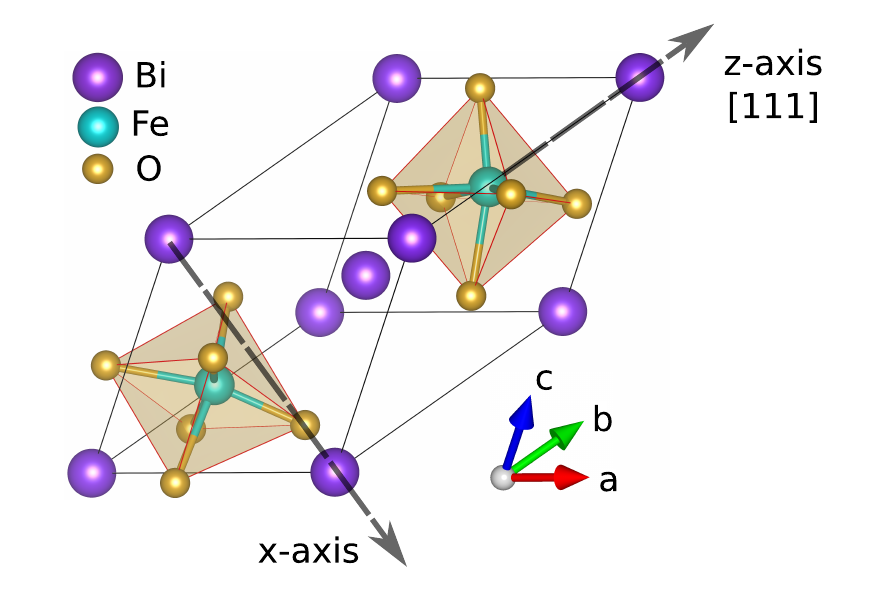}
    \caption{The primitive unit cell of BiFeO$_3$ used in the calculations.
    It includes two bismuth (purple) and iron (cyan) atoms each and six oxygen atoms (yellow).
    Additional atoms outside or on the edge of the unit cell are shown for better illustration of the geometry, \textit{e.g.}, to illustrate the distorted octahedron spanned by the six nearest-neighbor oxygen atoms of an iron atom (shaded planes).
    The coordinate system is chosen such that the $z$-axis coincides with the [111] crystal direction}
    \label{fig:structure}
\end{figure}

All calculations are done for a unit cell of BiFeO$_3$ comprising two formula units.
This allows a description of the antiferromagnetic order by allowing two adjacent Fe atoms to have opposite magnetization.
A two-formula-unit unit cell is also necessary for the description of the ferroelectric order and the associated structural distortion along the [111] crystal direction.
The ground state atomic structure is depicted in Fig.~\ref{fig:structure} and possesses \textit{{R3c}} space group symmetry.

The ground state Kohn-Sham DFT calculations were done with the \texttt{PWscf} code of the \texttt{QuantumESPRESSO}~\cite{hk-dft,ks-dft,Giannozzi2009,Andreussi2017-tq,Giannozzi2020-ha,pseudodojo_2018} suite.
The electronic wave functions and electronic charge density were expanded in a plane-wave basis set with an energy cutoff of 45~Ha and 180~Ha, respectively.
The interaction between the ionic core and the valence electrons is described by optimized, norm-conserving Vanderbilt (ONCV) pseudopotentials as provided by the \texttt{PseudoDOJO} repository~\cite{Hamann2013,pseudodojo_2018}.
We treat a total of 98 electrons per unit cell as valence electrons.
To obtain the ground state geometry via the minimization of the total energy, the electronic spins were restricted to be collinear.
For all subsequent calculations within the fixed ground-state geometry, however, this constraint was relaxed and the spin-orbit  interaction is taken into account through the use of fully relativistic pseudopotentials.
Wave vectors from the first Brillouin zone (BZ) are resolved by sampling the BZ with a uniform mesh of 6$\times$6$\times$6 points.
We compute a total of 600 Kohn-Sham quasi-particle states and energies that serve as the starting point for our $GW$ and $BSE$ calculations.

The $GW$ and BSE calculations are performed using the \texttt{Yambo} code~\cite{Marini2009,Sangalli2019}.
For the $GW$ calculations, we use the full set of 600 Kohn-Sham states in the calculation of the screened Coulomb interaction and for the correlation part of the electron self-energy.
We explicitly calculate the quasi-particle energies for the highest 26 occupied and lowest 30 unoccupied bands.

We studied the numerical convergence of the quasi-particle band gap with respect to several computational parameters in detail.
While the results of these convergence studies can be found in section 1 of the Supplemental Material (SM)~\cite{si}, we want to briefly comment on the convergence of the $G_0W_0$ case here.
In particular, we want to note that the convergence of the band gap on the $G_0W_0$-level with respect to the plane-wave energy cutoff used in the calculation of $W$ is very slow, see Fig.~S2 in the SM~\cite{si}.
Increasing this plane-wave energy cutoff from 6~Ha to 20~Ha leads to a shift of the $G_0W_0$ band gap by -0.2~eV, whereas the relative energies of the unoccupied states remain unaffected.
As the computational costs for the calculations with the higher plane-wave cutoffs become prohibitively demanding for the partially self-consistent ev$GW$ calculations, we use in those a 6~Ha plane-wave energy cutoff for the calculation of $W$ and apply a rigid shift of -0.2~eV to all quasi-particle energies for the unoccupied states afterwards.
The high plane-wave energy cutoff needed to accurately resolve the screened Coulomb potential $W$ points to a strong variation of the microscopic dielectric function on the atomistic scale related to a very spatially inhomogeneous electron density.

For the BSE calculations, we obtain the statically screened Coulomb interaction using only 200 Kohn-Sham states in the polarizability.
This is sufficient to obtain converged results, as the contribution of even higher energy bands is increasingly negligible at zero frequency.
We build the electron-hole interaction kernel using the 26 highest occupied and 30 lowest unoccupied Kohn-Sham bands, whose quasi-particle energies were obtained within the ev$GW$ framework.
We checked the convergence of the absorption spectrum obtained by solving the BSE with respect to the number of bands considered in the transition basis, see Fig.~S4 in the SM~\cite{si}.
To converge the optical absorption spectrum from the onset of the continuous spectrum onward, a rather modest plane-wave cutoff of 2.5~Ha for building the BSE interaction kernel is sufficient.
On the other hand, to converge the strongly bound excitations below the continuum and within the band gap, we used a comparatively much larger cutoff energy of 18~Ha to obtain converged results.
We attribute this physically to the attractive screened Coulomb interaction between the electron and the hole, which needs to bind two particles on small length scales that can only be resolved by plane wave with wave vectors of large magnitudes.
As shown in Fig.~S6 of the SM~\cite{si}, the lowest-energy spin-flip exciton (at an energy of 1.25 eV) is still not fully converged in our calculations due to its extremely strong localization (binding energy $>$3~eV).
However, as this is the only state that suffers from this problem and a further decrease of its energy will not change its nature and interpretation of being a strongly bound spin-flip exciton.
We extrapolate its energy in the limit of a complete basis set (1/ plane wave cutoff energy $\rightarrow$ 0) to a value of 0.82~eV. 
A summary of the convergence parameters used in the calculation is given is given in Tab.~S1 of the SM~\cite{si}.
Finally, for the DFT+U calculations, we use a $U$ value of 4~eV to penalize the delocalization of Fe~$3d$ electrons.

\section{Results}

\subsection{Electronic structure}

\begin{figure*}
    \centering
    \includegraphics[width=0.98\textwidth]{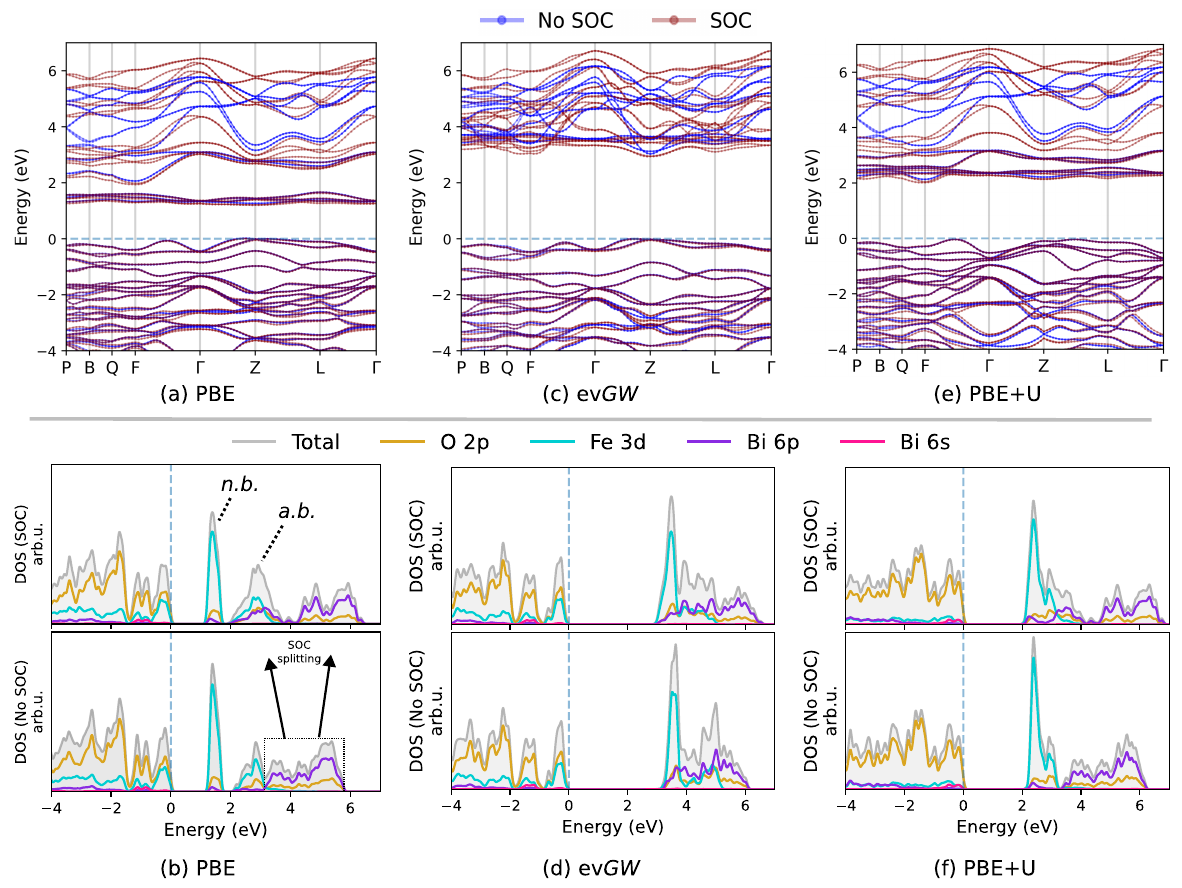}
    \caption{(a,c,e) Calculated electronic band structure of BiFeO$_3$ on the PBE (a), ev$GW$ (c), and PBE+U levels (e) along lines between high-symmetry points in the BZ.
    Blue lines represent the band structure without the inclusion of spin-orbit coupling, red lines with it.
    (b,d,f) Corresponding total and atom- and orbital-projected density of states.}
    \label{fig:e_str_dos}
\end{figure*}

\begin{figure}[]
    \centering
    \includegraphics[width=0.48\textwidth]{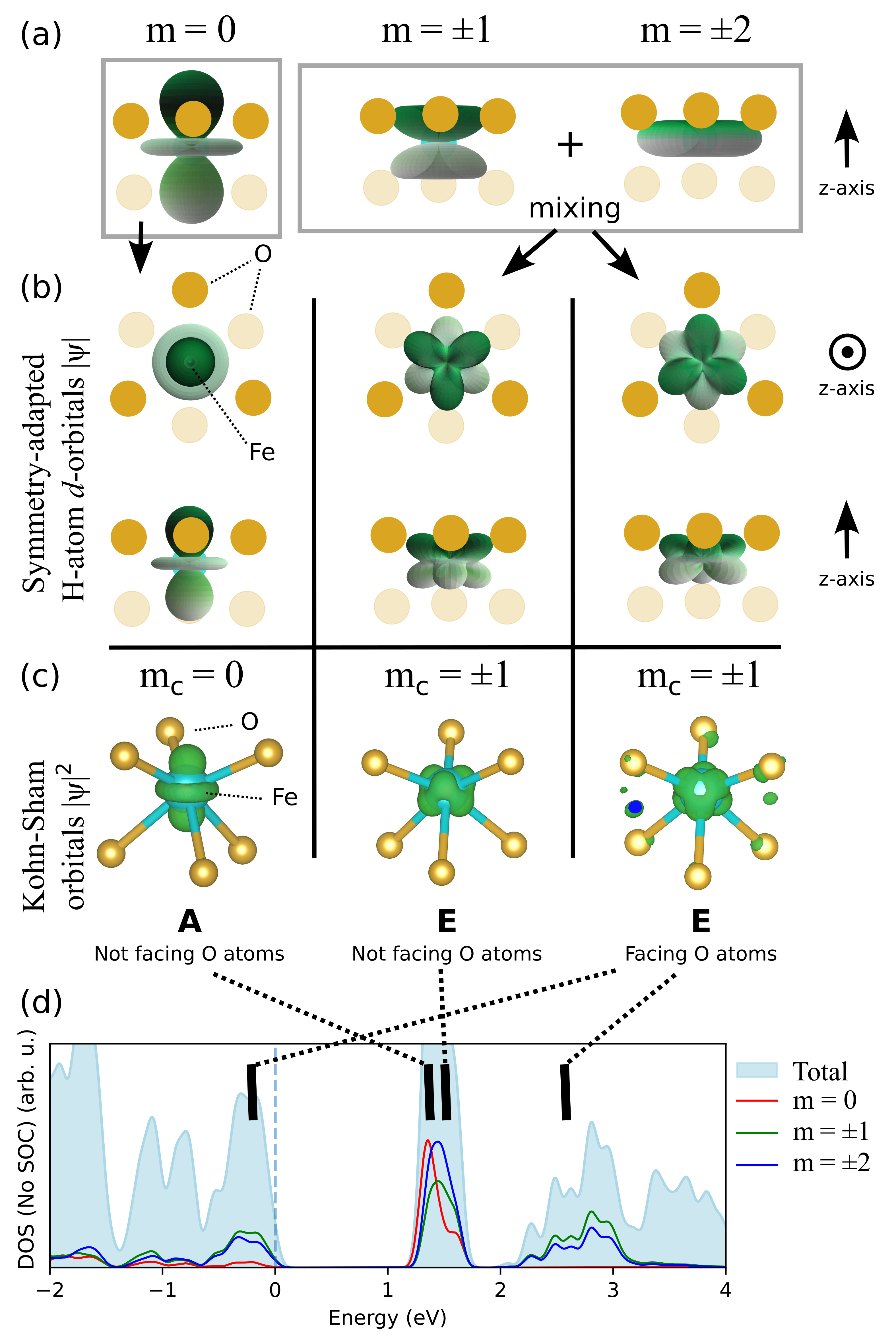}
    \caption{(a) H-atom $3d$ orbitals corresponding to a fixed absolute value of the azimuthal angular momentum quantum number. The golden points represent the six neighboring sites of the oxygen atoms.
    (b) Symmetry-adapted H-atom $3d$ orbitals with mixing angle as given in the text.
    (c) Kohn-Sham orbitals at $\mathbf{k}=\Gamma$ for selected states with significant Fe-$3d$ character.
    (d) Atomic-orbital projected density of states (without considering spin-orbit coupling).}
    \label{fig:sym_adapt_orbs}
\end{figure}

The electronic band structure computed on the level of DFT with the PBE functional without spin-orbit coupling (SOC) is shown in Fig.~\ref{fig:e_str_dos}a as blue lines.
The corresponding atom- and orbital-projected density of states (DOS) is shown in the lower panel of Fig.~\ref{fig:e_str_dos}d.
We note that the bands near the top of the valence band are composed of a mixture of Fe-$3d$ and O-$2p$ orbitals.
The mixing of orbitals of different atoms leads to some degree of delocalization of the electrons and is responsible for the slight dispersion of the energy levels.
It also has an impact on the interpretation of the optical spectrum of BiFeO$_3$, as any electronic transition involving states near the band gap necessarily involves the O-$2p$ orbitals and cannot be thought of as transitions occurring exclusively on an Fe site, although in the following we will adopt the wide-spread nomenclature of ``on-site Fe transitions'' for brevity.
The high-spin $3d^5$ electronic configuration of the Fe atoms in their 3$^+$ oxidation state leads to an opposite spin polarization on the two distinct Fe sites.
Beyond the valence band, the band structure features two distinct conduction bands separated by an energy gap of around 0.33~eV between each other and by a gap of 1.21~eV from the valence band.
The first, narrow conduction band is only weakly dispersive and can be associated with localized Fe-$3d$ states.
The second, wide conduction band is strongly dispersive and made up of almost equal mixtures of O-$2p$, Fe-$3d$, and Bi-$6p$ states.
As both of these conduction bands have significant Fe-$3d$ character, they can be interpreted in terms of two sets of unoccupied Fe-$3d$-states, split in energy due to the oxygen ligand field.
This is analogous to the splitting of the $d$-orbital states of an octahedrally coordinated Fe$^{3+}$ ion into a triplet $t_{2g}$-state and a doublet $e_g$-state.

In the actual BiFeO$_3$ crystal, however, the oxygen octahedra are deformed: the real geometry resembles two equilateral triangles of oxygen atoms of different side lengths and located at different distances from the central Fe atom.
The orientation of the two triangles of oxygen atoms is opposite up to a small additional twist angle of $\sim 1.27^{\circ}$, i.e., up to a translation in space and a stretch factor, the two O-triangles can be transformed into one another by a rotation about 1.27$^{\circ}$ around the $z$-axis .
This deformed octahedral geometry implies that the local symmetry around one Fe atom is reduced from the perfect $O_h$ octahedron group to the much smaller $C_3$ local point group symmetry.
As a consequence, the ``traditional'' $t_{2g}$-$e_g$ level splitting only holds approximately and the picture of the crystal field splitting in BiFeO$_3$ is better understood in a set of orbitals that are adapted to the actual symmetry.
For this, we first note that the local point group $C_3$, by virtue of being Abelian, possesses three irreducible, complex representations, $\Gamma_1 = A$, $\Gamma_2$, and $\Gamma_3 = \bar{\Gamma}_2$, all of which are necessarily one-dimensional~\cite{bradley1973}.
The most important feature and difference between these three representations is the behavior under 120$^{\circ}$ rotations around the $z$-axis.
$A$-states are invariant under rotations (crystal angular momentum quantum number $m_c=0$), while the $\Gamma_2$- ($\bar{\Gamma}_2$-) states pick up a phase factor of $\mathrm{e}^{+i 2 \pi/3}$ ($\mathrm{e}^{-i 2 \pi/3}$), corresponding to $m_c=+1$ ($m_c=-1$).
Most importantly, the one-dimensionality of the representations ensures that there are no exact degeneracies on-site.
Physically, this is a consequence of the geometrical fact that, from the point of view of the central Fe atom, there is a difference in the environment between moving in a clockwise or in a counter-clockwise way.
In other words, the finite oxygen-triangle twist angle identifies a preferred rotation orientation.
As a consequence, states with positive or negative angular momentum split in energy.
We note however, that if the twist angle between the two oxygen triangles is exactly $0^{\circ}$ or $180^{\circ}$, there is no physical difference between the environments in the clockwise or counter-clockwise direction and the two states of finite, opposite angular momentum are degenerate.
From a group theoretical point of view, this has its roots in the additional vertical mirror plane symmetry in these high-symmetry scenarios, i.e., the local symmetry group is enlarged to $C_{3v}$.
This group allows, among the finite angular momentum representations, only a doubly degenerate $E$ representation and thus the $m_c=+1$ and $m_c=-1$ are exactly degenerate in this case.
In the actual BiFeO$_3$ geometry at a non-symmetric oxygen triangle twist angle, however, this vertical mirror symmetry is slightly broken and the splitting between the $m_c=+1$ and $m_c=-1$ states technically becomes finite, even though at a twist angle of only $\sim 1.27^{\circ}$, it is below the computational resolution.

As discussed above, group theory predicts for the unoccupied $3d$-states localized on one Fe atom that they are either singly degenerate with $m_c=0$ total crystal angular momentum quantum number or that they are almost doubly degenerate with $m_c=\pm 1$, only split by the small twist angle between the oxygen triangles.
It does not, however, predict the relative location of the the $m_c=0$ and $m_c=\pm1$ states on the energy scale.
To understand this, we analyze the electronic density of states in terms of the contribution of symmetry-adapted Fe-$3d$-orbitals.
The standard $t_{2g}$-$e_g$ split is traditionally understood in terms of the difference in Coulomb repulsion for electrons in the set of the $d_{xz}$-, $d_{yz}$-, and $d_{xy}$-orbitals on the one hand and electrons in the $d_{x^2-y^2}$- and $d_{z^2}$-orbitals on the other hand.
In the actual $C_3$-symmetric oxygen environment of one Fe atom in BiFe$O_3$, however, this set of orbitals is no longer suitable to analyze the level splitting.\par 

Instead, we propose a new, symmetry-adapted set of Fe-$3d$-orbitals as a more suitable basis in which to analyze the electronic density of states.
From the symmetry considerations discussed above, we first group the set of five Fe-$3d$-orbitals into three sets, according to their commonality with regard to rotations around the $z$-axis by $120^{\circ}$:
(i) the trivially transforming $d_0 = d_{z^2}$-orbital with angular momentum quantum number $m=0$,
(ii) the $d_{+1} \propto (d_{xy} + i d_{yz})$- and $d_{-2} \propto (d_{x^2-y^2} - i d_{xy})$-orbitals, which have $m=+1$ and $m=-2$, respectively, i.e., the same crystal angular momentum of $m_c = m \text{ mod } 3 = +1$,
and (iii) the $d_{-1} \propto (d_{xy} - i d_{yz})$- and $d_{+2} \propto (d_{x^2-y^2} + i d_{xy})$-orbitals, which have $m=-1$ and $m=+2$, respectively, i.e., $m_c = 2 \text{ mod } 3 = -1$.
Orbitals from different sets cannot mix with each other by symmetry, so that only the two states in the $m_c=+1$ and the $m_c=-1$ sets can mix among themselves.
We illustrate the $d$-orbitals of definite angular momentum in Fig.~\ref{fig:sym_adapt_orbs}a.
With a suitable choice of phase of the constituent $d$-states, we can write the five symmetry-adapted Fe $3d$-orbitals in terms of two real mixing angles $\theta_{+1}$ and $\theta_{-1}$:
\begin{equation}
\begin{split}
      (i) \ & | m_c=0; \text{n.-b.}\rangle = |d_0 \rangle, \\
     (ii) \ & | m_c=+1; \text{n.-b.} \rangle = \cos \theta_{+1} | d_{+1} \rangle + \sin \theta_{+1} | d_{-2} \rangle, \\
            & | m_c=+1; \text{a.-b.} \rangle = -\sin \theta_{+1} | d_{+1} \rangle + \cos \theta_{+1} | d_{-2} \rangle, \\
    (iii) \ & | m_c=-1; \text{n.-b.} \rangle = \cos \theta_{-1} | d_{-1} \rangle + \sin \theta_{-1} | d_{+2} \rangle, \\
            & | m_c=-1; \text{a.-b.} \rangle = -\sin \theta_{-1} | d_{-1} \rangle + \cos \theta_{-1} | d_{+2} \rangle,
\end{split}
\label{eq:sym_orbs}
\end{equation}
where the labels ``n.-b.'' and ``a.-b.'' stand for ``non-bonding'' and ``anti-bonding'' respectively and are used in the $m_c=\pm1$ case to distinguish between the states and in the $m_c=0$ case for clarity.
Note that in the context of the band structure of the crystal as a whole, the states and mixing angles become functions of the wavevector $\mathbf{k}$.
However, as the lowest conduction bands are not very dispersive due to the localization of the corresponding states, we determine the mixing angles for the case $\mathbf{k}=\Gamma$ only.
To obtain the mixing angle, we compute the overlaps of the Kohn-Sham states $|\mathbf{k}=\Gamma,n\rangle$ with the five local Fe-$3d$-orbitals.
In the first conduction band, there are six Kohn-Sham states at $\mathbf{k}=\Gamma$, three localized to the largest extend on each of the two Fe atoms.
We compute the mixing angles from those two states that have the smallest contributions of non-Fe-$3d$-orbitals.
Numerically, we find mixing angles of $\theta_{+1}=48.37^{\circ}$ and $\theta_{-1}=48.29^{\circ}$.
These indicate that the small O-triangle twist angle only leads to a minimal difference in the orbital mixing of states of positive or negative crystal angular momentum, as expected from the absence of any observable splitting between the $m_c=+1$ and $m_c=-1$ states.
More importantly, it also shows that the ``standard'' basis for octahedral crystal field splitting of $d_{xz}, d_{yz}, d_{xy}$, etc. breaks down completely, as there is a large degree of mixing between these orbitals.

In Fig.~\ref{fig:sym_adapt_orbs}b we show the set of five symmetry-adapted $d$-orbitals.
The two states $|m_c=\pm1,\text{n.-b.}\rangle$ together with the state $|m_c=0\rangle$ form the first conduction band.
All three of these orbitals are oriented such that they point away from the oxygen atoms.
In quantum chemistry terms, these states therefore have \emph{non-bonding} character.
By contrast, the orbitals corresponding to the states $|m_c=\pm1,\text{a.-b.}\rangle$ are pointing toward the oxygen atoms.
These states therefore participate in the \emph{bonding} and \emph{anti-bonding} bands that form the top of the valence and bottom of the second conduction band, respectively.
The orientation of the orbitals in the respective bands is furthermore confirmed by a look at the probability density of the corresponding Kohn-Sham states, see Fig.~\ref{fig:sym_adapt_orbs}c.
For both the first conduction band and the bottom of the second conduction band, the shape of the \textit{ab initio} probability density is in very good agreement with the prediction of the group theoretical analysis.

\begin{figure*}
    \centering
    \includegraphics[width=0.98\textwidth]{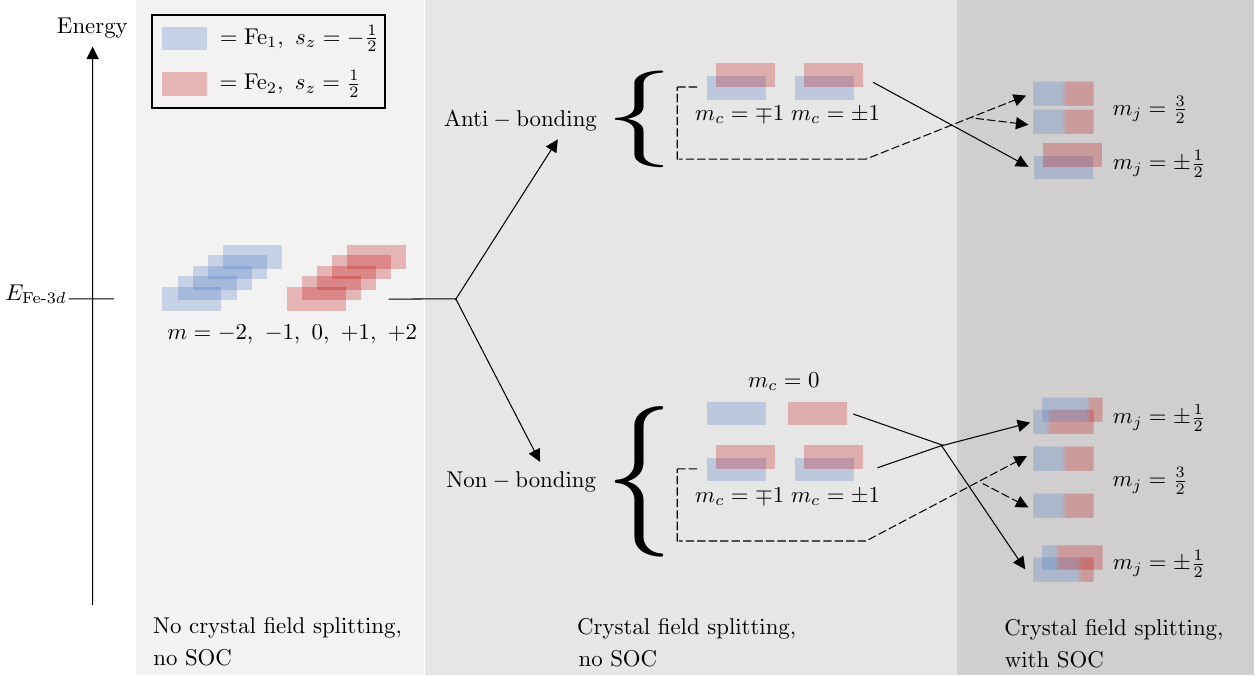}
    \caption{Qualitative sketch of the crystal field splitting of the \emph{unoccupied} Fe-$3d$ states with and without spin-orbit coupling.
    Each rectangle represents one electronic state.
    Rectangles stacked into the plane denote \emph{exactly} degenerate states by symmetry, horizontally aligned states denote \emph{de facto} degenerate states.
    The labels indicate (crystal) angular momentum number.
    In case of a doubly degenerate state, the upper sign corresponds to the state at the front of the stack.
    The color refers to the qualitative character of the state: blue for the Fe atom with spin down in the unoccupied state, red for the Fe site with unoccupied spin-up states.
    A rectangle of mixed color indicates a perfectly site-mixed state.
    A rectangle of predominantly one color indicates a state with a dominant contribution of one Fe site and a minority contribution of the other.
    The level splittings are \emph{not} to scale.}
    \label{fig:crystal-field-splitting-scheme}
\end{figure*}

So far, we have treated the $3d$-states of the two inequivalent Fe atoms as independent.
Within the context of the $R3c$ space group of the crystal, however, the two sets of five symmetry-adapted Fe-$3d$-orbitals transform according to the irreducible representations of the $R3c$ space group.
For instance, at $\mathbf{k}=\Gamma$, the electronic states transform as one of the three irreducible representations of $R3c$: $A_1$, $A_2$, or $E$, the latter of which is doubly degenerate.
Since only the $m_c=\pm1$ states transform non-trivially under rotations, the $m_c=+1$ and $m_c=-1$ states of both the \emph{non-bonding} and \emph{anti-bonding} type form an $E$ doublet and are hence exactly degenerate.
We note that this degeneracy could be lifted by chemically substituting one of the two Fe sites with a different magnetic atom.
This could be desirable if an energy gap between the states of positive and negative chirality is desired for an application.
Finally, the two $m_c=0$ states, one localized on one Fe site, the other on the second, are coupled, with the symmetric linear combination being the $A_1$ state in the context of the space group symmetry and the anti-symmetric linear combination transforming as an $A_2$ state.
However, as the two $m=0$ states are localized on different Fe sites, they have opposite spin character and as a result cannot couple directly in the absence of spin-orbit interaction.
In consequence, the $A_1$ and $A_2$ states remain degenerate.
We summarise the crystal field splitting in Fig.~\ref{fig:crystal-field-splitting-scheme}, where we also indicate the site and spin character of each state.

Upon inclusion of the spin-orbit coupling (SOC), the bandstructure undergoes significant changes, starting at the second conduction band (see red lines in Fig.~\ref{fig:e_str_dos}a and top panel in Fig.~\ref{fig:e_str_dos}b). 
The SOC leads to a large splitting most clearly seen in the DOS, where the accumulation of state density around 4~eV is redistributed into two separate peaks centered around 3 and 5~eV.
Based on the changes in the positions of the local DOS maxima, the average magnitude of the SOC splitting can be estimated to be around 0.73 eV.
As seen in the orbital-resolved DOS, the SOC splitting is most significant for conduction bands with a significant contribution of the Bi-$6p$-orbitals.
We also note that the strong SOC leads to a reordering of the subbands, as it pushes the bands with higher Bi-$6p$ admixture below some of the subbands with less contribution from Bi orbitals.

The effect of SOC on the crystal-field-split Fe-$3d$ orbitals, which compose the first conduction band and the bottom of the second, can be analyzed by considering the spin degree of freedom of these unoccupied states and the total crystal angular momentum $m_j$ it gives rise to.
Due to the antiferromagnetic ground state order, the unoccupied $3d$-states on the two distinct Fe sites have opposite $z$ components of the spin angular momentum.
Between the two Fe sites, the five symmetry-adapted orbitals constructed in Eq.~\eqref{eq:sym_orbs} for each of the Fe atoms then form a set of ten basis functions with three inequivalent total crystal angular momenta $m_j=-1/2$, $m_j=+1/2$, and $m_j=+3/2$, where it should be noted that due to the $120^{\circ}$-symmetry, a total angular momentum of $-3/2$ is equivalant to $+3/2=-3/2+3$.
Grouped by total crystal angular momentum number, the ten basis states are given by
\begin{widetext}
\begin{eqnarray}
       (i) \ & | m_j=-\tfrac12; \text{n.-b.}; \mathrm{Fe}_1 \rangle & \ \equiv \ | m_c=0; \text{n.-b.}; m_s=-\tfrac12 (\mathrm{Fe}_1) \rangle \label{eq:states_SOC} \\
             & | m_j=-\tfrac12; \text{n.-b.}; \mathrm{Fe}_2 \rangle & \ \equiv \ | m_c=-1; \text{n.-b.}; m_s=+\tfrac12 (\mathrm{Fe}_2) \rangle \nonumber \\ 
             & | m_j=-\tfrac12; \text{a.-b.}; \mathrm{Fe}_2 \rangle & \ \equiv \ | m_c=-1; \text{a.-b.}; m_s=+\tfrac12 (\mathrm{Fe}_2) \rangle \nonumber \\
      (ii) \ & | m_j=\tfrac12; \text{n.-b.}; \mathrm{Fe}_2 \rangle & \ \equiv \ | m_c=0; \text{n.-b.}; m_s=+\tfrac12 (\mathrm{Fe}_2) \rangle \nonumber \\
             & | m_j=\tfrac12; \text{n.-b.}; \mathrm{Fe}_1 \rangle & \ \equiv \ | m_c=+1; \text{n.-b.}; m_s=-\tfrac12 (\mathrm{Fe}_1) \rangle \nonumber \\
             & | m_j=\tfrac12; \text{a.-b.}; \mathrm{Fe}_1 \rangle & \ \equiv \ | m_c=+1; \text{a.-b.}; m_s=-\tfrac12 (\mathrm{Fe}_1) \rangle \nonumber \\
     (iii) \ & | m_j=\tfrac32; \text{n.-b.};  \mathrm{Fe}_1 \rangle & \ \equiv \ | m_c=-1; \text{n.-b.}; m_s=-\tfrac12 (\mathrm{Fe}_1) \rangle \nonumber \\
             & | m_j=\tfrac32; \text{n.-b.}; \mathrm{Fe}_2 \rangle & \ \equiv \ | m_c=+1; \text{n.-b.}; m_s=+\tfrac12 (\mathrm{Fe}_2) \rangle \nonumber \\ 
             & | m_j=\tfrac32; \text{a.-b.}; \mathrm{Fe}_1 \rangle & \ \equiv \ | m_c=-1; \text{a.-b.}; m_s=-\tfrac12 (\mathrm{Fe}_1) \rangle \nonumber \\ 
             & | m_j=\tfrac32; \text{a.-b.}; \mathrm{Fe}_2 \rangle & \ \equiv \ | m_c=+1; \text{a.-b.}; m_s=+\tfrac12 (\mathrm{Fe}_2) \rangle, \nonumber
\end{eqnarray}
\end{widetext}
where the $m_c=\pm1$ states on the right hand side are understood to be the states from Eq.~\eqref{eq:sym_orbs}.
The states within each group are allowed to couple and mix, while states from groups with different $m_j$ do not couple.
As the \emph{non-bonding} states and the \emph{anti-bonding} ones are separated by a sizable energy gap, the SOC can be considered to merely result in a small shift of the anti-bonding state, while resulting in a sizable repulsion of the two non-bonding states of the same $m_j=\pm 1/2$.
As a result, our symmetry analysis predicts two split non-bonding states and one slightly shifted anti-bonding state for each of the $m_j =+1/2$ and $m_j=-1/2$ group of states.
Due to the coupling between the two non-bonding states of the same $m_j$, which are localized on different Fe sites, the spin-orbit-coupled states acquire a degree of delocalization due to the admixture of the state localized on the opposite Fe site.
However, as the two coupled non-bonding states are already split in energy before the SOC (cf. Fig.~\ref{fig:crystal-field-splitting-scheme}), the coupled states will have a dominant component localized on one of the Fe sites and a minority component localized on the other.
In Fig.~\ref{fig:crystal-field-splitting-scheme}, this is qualitatively indicated by the coloring of the rectangle representing each state.
As for the case without SOC, the actual $R3c$ symmetry of the crystal ensures that each of the three states in a $m_j=\pm1/2$ group forms a degenerate doublet with the corresponding state of opposite $m_j$.
Finally, the four $m_j = 3/2 \equiv -3/2$ states split pairwise into four non-degenerate states, each with a 1:1 mixture of the two Fe sites, as the pairs of states with $m_c=\pm1$ in the non-SOC case are exactly degenerate due to the crystal symmetry.
The results of this symmetry analysis of the lowest-energy excited states of BiFeO$_3$ are confirmed by our DFT results.

Beyond density functional theory, on the ev$GW$~level, both the valence and conduction bands change noticeably, as shown in Fig.~\ref{fig:e_str_dos}c.
In the valence band, the bonding O-$2p$$-$Fe-$3d$ set of states at the top of the valence band form a separate band that is split off by an energy gap.
A comparison of the DFT- and ev$GW$-level DOS (Fig.~\ref{fig:e_str_dos}d) reveals that the split-off valence subband consists of the bonding O-$2p$$-$Fe-$3d$ hybridized orbitals, whereas the top of the lower valence subband consists almost entirely of the O-$2p$ orbitals that bind the oxygen atoms.
The splitting of the valence band can thus be traced back to the overscreening of the electron-electron interaction in the Fe-$3d$ admixture of the bonding orbitals.
The strong localization of electrons in these orbitals and the resulting short length scale over which their mutual Coulomb interaction takes place require a more accurate treatment of the dielectric function than what is offered by the semi-local mean-field approach of DFT.
The screened Coulomb interaction used in the ev$GW$ approach thus allows a better distinguishing of the electron-electron interaction in the O-Fe bonds and in the O-exclusive bands.
Beyond the valence band, the ev$GW$ bandstructure sees the localized Fe-$3d$ band merge with the rest of the conduction band for similar reasons. 
The rest of the conduction band is squeezed due to the lower energy levels acquiring a larger correction, as is well-known from quantum mechanical second-order perturbation theory.

By comparison, a single-shot $G_0W_0$ calculation on top of the DFT ground state opens up the DFT band gap by only 0.67 eV and yields an (absolute) direct gap of 1.89 eV.
As has been demonstrated in other cases, the single-shot $G_0W_0$ method often underestimates the band gap, as it depends strongly on DFT-level quantities.
By contrast, the ev$GW$ results are more reliable, as the dependence on the DFT-level electron energies is diminished in the partially self-consistent cycle.
As shown in Fig.~S3 of the SM~\cite{si}, 4-5 iterations in the eigenvalue self-consistency cycle are needed to achieve convergence of the quasi-particle eigenvalues. 

\begin{figure}[]
    \centering
    \includegraphics[width=0.48\textwidth]{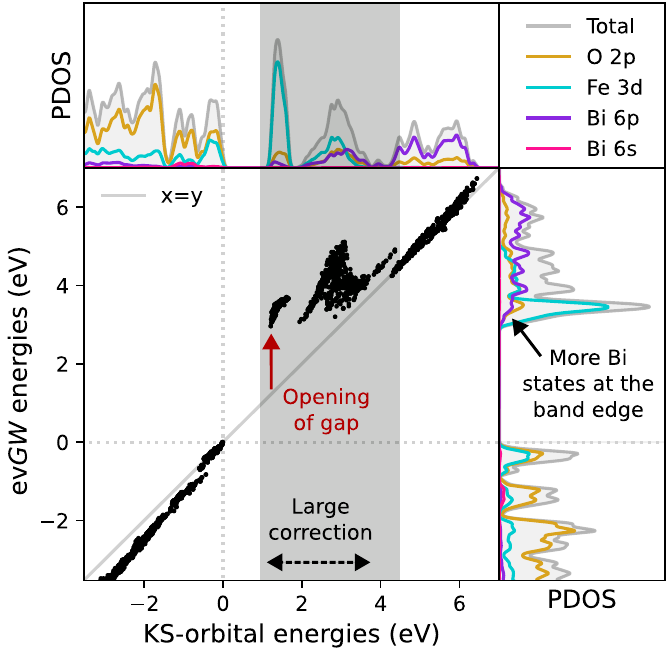}
    \caption{Kohn-Sham electron energies versus ev$GW$ quasi-particle energies with corresponding densities of states.}
    \label{fig:gw_conv}
\end{figure}

In Fig.~\ref{fig:gw_conv} we compare the  DFT-level energies of the Kohn-Sham states to the ev$GW$ quasi-particle energies and correlate it with the orbital character of the states by way of the corresponding partial density of states (PDOS).
As expected, the $GW$ corrections for states with significant Fe-$3d$ character are particularly large, owing to the typical over-delocalization of $d$ electrons in semi-local DFT.
In consequence, the correction is largest for the non-bonding states of the first conduction band, as they are almost exclusively formed by the Fe-$3d$ orbitals. 
The second largest correction in energy is experienced by the non-bonding states at the bottom of the second conduction band, as they are predominantly of Fe-$3d$ character as well.
The net result of the $GW$ correction is that the gap between the first and second conduction band is reduced significantly, to the point that the non-bonding and anti-bonding states merge to some extent in the ev$GW$-level PDOS.

This has a profound impact on the bandstructure upon the inclusion of spin-orbit coupling, as the non-bonding and anti-bonding states of each group in Eq.~\eqref{eq:states_SOC} now strongly interact, contrary to the case without SOC.
As a consequence, the SOC-induced splitting of the non-bonding and anti-bonding states is much larger, as can be seen, for example, in the PDOS in Fig.~\ref{fig:e_str_dos}e.
The edge of the conduction band in the SOC case (top panel) acquires a notable shoulder compared to the case without SOC (bottom panel).
This shoulder can be traced back to the interaction between the non-bonding and anti-bonding states, which we confirmed by checking the orbital composition of the KS states of lowest quasi-particle energy.
An observable consequence of this is the lowering of the quasi-particle gap by 0.1~eV.

Lastly, we compare the band structure and PDOS of our ev$GW$-level calculation to their counterparts on the level of PBE+U shown in Fig.~\ref{fig:e_str_dos}e and f.
Here, the inclusion of the Hubbard-$U$ term leads to the localized, non-bonding Fe-$3d$ orbitals being pushed into the second conduction band, which results in a quasi-particle gap of 2.02~eV that is also moved from the vicinity of the $Z$ point to the vicinity of the $F$~point, as already noted in earlier studies~\cite{clark_energy_2009,shenton_effects_2017,Ghosal2022-tc,Neaton2005}.
Upon the consideration of spin-orbit coupling, we observe a broadening of the conduction band edge, which, however, is less pronounced than in the ev$GW$ case, possibly due to the bigger energetic separation of non-bonding and anti-bonding states.
Lastly, the $U$ term also leads to the almost complete removal of the Fe-$3d$ content from the valence band.

Quantitatively, the ev$GW$ approach yields by far the best prediction for the quasi-particle band gap (2.76~eV vs. 2.67~eV as extracted from optical experiments~\cite{sando_revisiting_2018}).
By comparison, see Tab.~\ref{tab:bandgaps}, the quasi-particle band gap is significantly underestimated in the non-self-consistent $G_0W_0$ scheme (1.89~eV) and the semi-self-consistent ev$GW_0$ scheme (2.26~eV), in which the screened Coulomb potential $W$ is not updated iteratively once calculated with the DFT-level electron energies and wave functions, but the one-electron Green's function $G$ is through updating the electron energies.
The PBE+U band gap of 2.20 to 2.26~eV is also much smaller than is reported experimentally and is almost independent of the choice of $U$. 

Overall, we conclude from this that in terms of $GW$ ``flavor'', for BiFeO$_3$, a partially self-consistent scheme is necessary to arrive at excitation energies in agreement with experiment.
The PBE+U approach suffers from the two shortcomings that $U$ is a tunable parameter and that it is only applied to the Fe-$3d$ states, while all other states are treated on the level of semi-local DFT.
The fact that $U$ is tunable makes the method less predictive while at the same time, if $U$ varied on a reasonable scale, the experimentally observed quasi-particle band gap cannot be reproduced.
For instance, a change in $U$ from 4~eV to 6~eV merely changes the quasi-particle gap from 2.20 to 2.26~eV (see  Tab.~\ref{tab:bandgaps}), \textit{i.e.}, still much smaller than observed in experiment.
The physical reason for this can be found in the fact that excited states with excitation energy of more than 2~eV are composed to a significant part of O-$2p$ and Bi-$6p$ states (see Fig.~\ref{fig:e_str_dos}f), which are not affected by the $U$ interaction term.
In a recent study, Ghosal \textit{et al.} attempted to address this particular problem by including two separate $U$ parameters for the Fe-$3d$ and O-$2p$ orbitals.
This, however, adds yet another tunable parameter to the description, while our parameter-free ev$GW$ approach already yields a satisfactory agreement with experiment for the band gap.
For completeness, we also computed the band gap using the Gau-PBE hybrid functional, see Tab.~\ref{tab:bandgaps}.
Despite its large computational cost, this method results in an overestimation of the band gap by 0.34~eV, and an inaccuracy of a similar magnitude as that of PBE+U. \par 

\begin{table}[hb!]
\centering
\begin{tabular}{ccc}
\hline \hline
\textbf{Method} & \textbf{Direct gap} (eV) & \textbf{Indirect gap} (eV) \\ \hline \hline
PBE       & 1.22 (Z$^{*}$-Z$^{*}$) & 1.21 (Z$^{*}$-Z) \\
PBE+U ($U$=4~eV)     & 2.20 (F$^*$-F$^{*}$) & 2.02 (F$^*$-F$^{*}$) \\
PBE+U ($U$=6~eV)     & 2.26 (F$^*$-F$^{*}$) & 2.01 (F$^*$-F$^{*}$) \\
Gau-PBE   & 3.01 (Z$^{*}$-Z$^{*}$) & 3.01 (F$^*$-Z$^{*}$) \\ 
$G_0W_0$  & 1.89 (Z$^{*}$-Z$^{*}$) & 1.88 (Z$^{*}$-Z) \\
ev$GW_0$  & 2.26 (Z$^{*}$-Z$^{*}$) & 2.24 (Z$^{*}$-Z$^{*}$) \\
ev$GW$    & 2.76 (Z$^{*}$-Z$^{*}$) & 2.74 (Z$^{*}$-Z$^{*}$) \\
Expt.~\cite{sando_revisiting_2018}     & 2.67 & -    \\
\hline \hline
\end{tabular}
\caption{Electronic band gaps of BiFeO$_3$ calculated on different levels of theory.
    The superscript $*$ on a high-symmetry point in the first Brillouin zone indicates that a point in close vicinity of that high-symmetry point is meant instead of the point itself.} \label{tab:bandgaps}
\end{table}

\subsection{Optical absorption spectrum}

\begin{figure}[h!]
    \centering
    \includegraphics[width=0.485\textwidth]{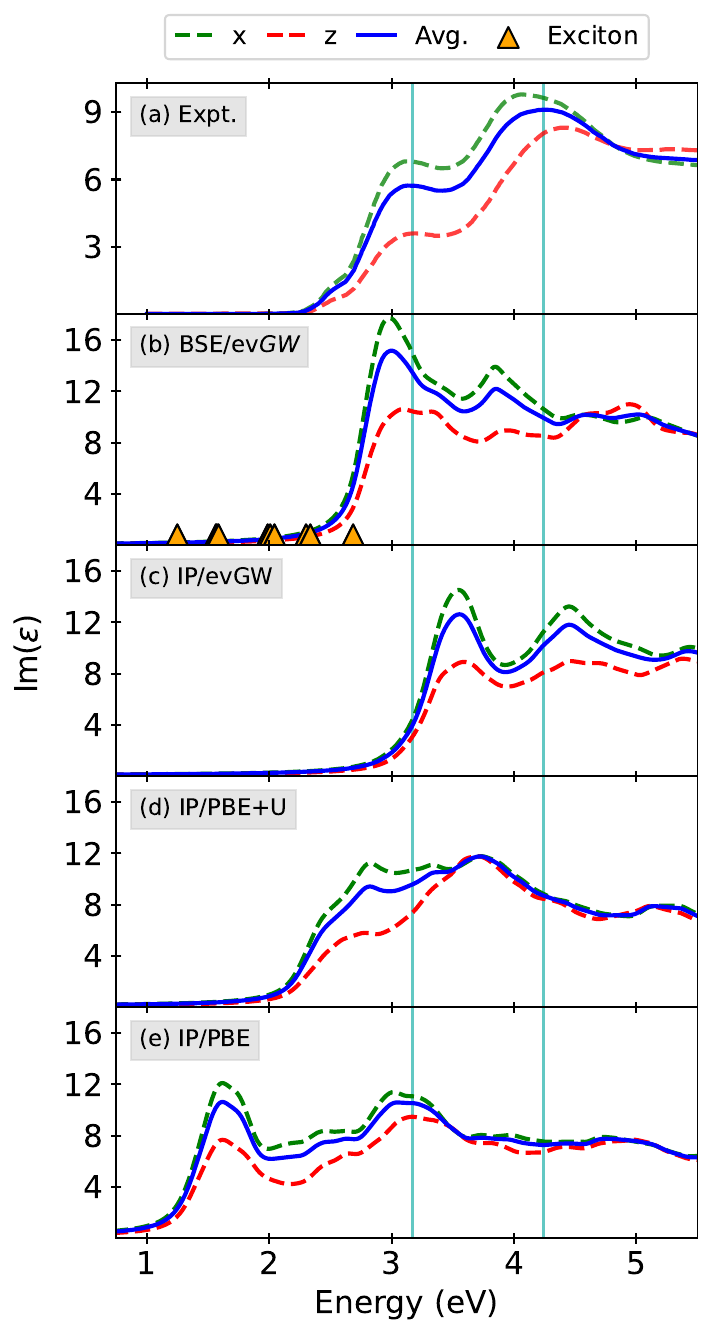}
    \caption{Imaginary part of the dielectric function from experiment and calculated on different levels of theory and for different light polarization: x (red, dashed line), z (green, dashed line), and averaged over x-, y-, and z-polarizations (blue, full line).
    (a) Experimental spectra, taken from Ref.~\cite{choi_optical_2011}.
    (b,c,d) Calculated spectra on different levels of theory: (b) BSE@ev$GW$, (c) independent particles on the level of ev$GW$, (d) independent particles on the level of PBE+U, (d) independent particles on the level of PBE.
    The teal lines mark the energies at which the two local maxima in the polarization-averaged experimental spectrum occur.
    The yellow triangle mark the energies of individual excitons that contribute to the dielectric function on the BSE@ev$GW$-level.}
    \label{fig:optical_spectra}
\end{figure}

We compare the optical absorption spectrum calculated with different computational methods and for different light polarization directions with the experimentally observed spectrum for a single crystal of BiFeO$_3$~\cite{choi_optical_2011} in Fig.~\ref{fig:optical_spectra}.
We compare the spectrum for light polarized along the x-axis ([110] crystal direction; red, dashed lines in Fig.~\ref{fig:optical_spectra}), for light polarized along the z-axis ([111] crystal direction; green, dashed lines), and for unpolarized light (average of x-, y-, z-polarization; blue lines).
The absorption spectrum features the well-known birefringence, which can be traced back to the difference in oscillator strength for $A_1$- and $A_2$-excitons, which couple to z-polarized light, on the one hand, and the $E$-excitons, which couple to x- or y-polarized light, on the other.
The polarization-averaged experimental spectrum features two local maxima at energies of 3.17~eV and 4.24~eV, which are highlighted by vertical teal lines for easier comparison to the calculated spectra.

We compare the experimental spectrum to the absorption spectrum calculated on four different levels of theory: solving the BSE with quasi-particle energies from ev$GW$ (b), independent particles (IPs) with the quasi-particle energies from ev$GW$ (c), IPs from PBU+U (d), and IPs from PBE (e).
The IP spectrum calculated on the level of DFT/PBE (panel e) qualitatively capture the two-peak structure of the experimental spectra and the birefringent behavior of both absorption maxima.
However, as expected, it fails to quantitatively describe both the absolute and the relative positions of the two maxima.
The IP spectrum on the PBE+U-level (panel d) is improved in this regard, as the relative position of the two maxima is approximately correct.
However, the overall shape of the absorption spectrum strongly diverts from the experimental shape, as it does not feature two clearly visible maxima anymore, but is rather smooth and more uniform.
In addition, the birefringence of the second absorption maximum is non-existent, in stark contrast to the experimental observation.
On the ev$GW$-level (panel c), the two-peak structure of the absorption spectrum is predicted reasonably well, as is the birefringent behavior.
This suggests that an accurate knowledge of the quasi-particle energies is the most crucial ingredient to capture the most prominent feature of the absorption spectrum of BiFeO$_3$.
Finally, the inclusion of the attractive electron-hole attraction in the most sophisticated, BSE@ev$GW$ approach redshifts the spectrum and brings it in relatively good agreement with experiment, both in terms of the presence of two clear local maxima and in terms of their absolute and relative positions.
One persisting shortcoming of both the BSE@ev$GW$ and IP@ev$GW$ methods, however, is their disability to correctly capture the intensity ration of the two absorption maxima.
We ascribe this to the use of the electron and hole orbitals from DFT/PBE in place of the quasi-particle orbitals in the calculation of optical transition matrix elements.
We note that neither of the employed theoretical approaches is able to fully reproduce the leading shoulder in the experimental spectrum at an energy of around 2.5~eV, which may be consistent with it having been argued to be due to defect states.  
In Fig.~1c of our companion paper~\cite{aseem_sven_prl}, the BSE/ev$GW$-level spectrum is plotted with very small excitonic broadening parameter and it shows a complex structure at the absorption onset with exciton states between 2.5~eV and an absorption onset at 2.76~eV.
These states can possibly contribute to the shoulder-like feature at $\sim$~2.5 eV.
However, a realistic description of the optical features at the onset would require a calculation of the state-specific broadenings, which is beyond the scope of this work.  A recent computational study of the optical spectra of BiFeO$_3$ by K\"{o}rbel claims that the shoulder-like feature has an intrinsic character~\cite{korbel2023}, however without further elaboration on the origin of the feature and character of the associated electronic transitions.  \par 

Our BSE calculations also predict optically weak but strongly bound exciton states below the onset of the absorption spectrum.
The positions of these states on the energy axis are marked in Fig.~\ref{fig:optical_spectra}b with triangular markers.
These exciton states correspond to \emph{spin-flip} excitations, involving electronic transitions from hybridised Fe-$3d$$-$O-$2p$ orbitals to Fe-$3d$ states and are effectively localized on only one of the two Fe sites within the unit cell.
While these excitations are spin-forbidden in the absences of spin-orbit coupling, the SOC enables their coupling to light.
In our companion work~\cite{aseem_sven_prl}, we focus on their coupling to circularly polarized light and also take a look at \emph{spin-conserving} transitions, which involve the hopping of an electron from an Fe-O bond to the nearest \emph{inequivalent} Fe center.
These excitations contribute to the absorption spectrum both above and below the absorption onset and are only weakly bound and thus delocalized in space (see Figs.~1 and 2 of Ref.~\cite{aseem_sven_prl}).

\begin{figure}[]
    \centering
    \includegraphics[width=0.485\textwidth]{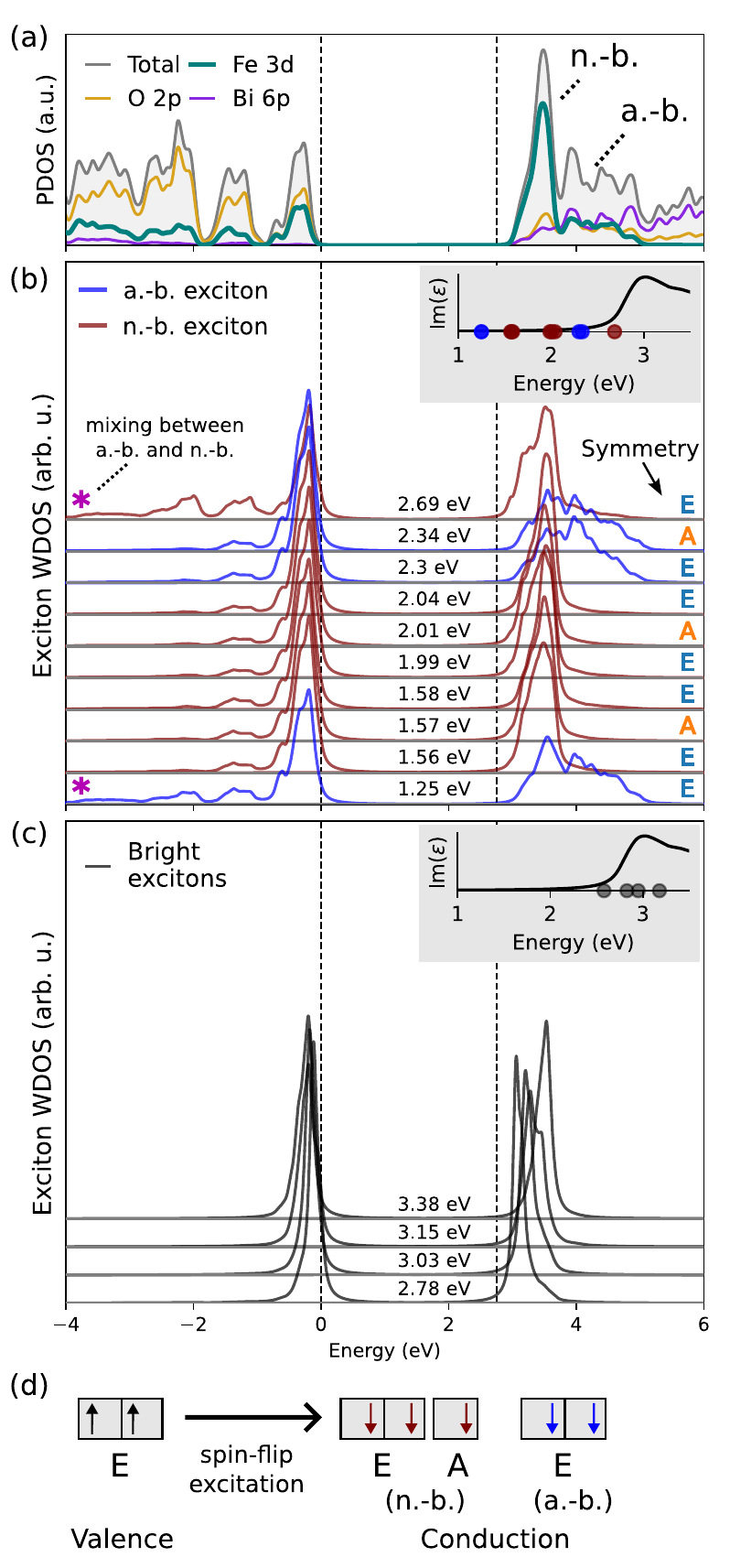}
    \caption{Comparison of exciton-weighted DOS of the electron and hole components of strongly bound spin-flip excitons (b) and the brightest spin-conserving excitons (c) with the ev$GW$ PDOS (a). The spin-flip excitons are categorized as non-bonding (n.-b.) and anti-bonding (a.-b.) excitons, according to the character the electron component of the exciton.
    Inset in (b) and (c): corresponding positions of spin-flip and spin-conserving excitons, respectively, in the optical absorption spectrum.
    (d) Schematic of the spin-flip transitions.}
    \label{fig:t2g_eg_excitons}
\end{figure}

In order to further analyze the optically active excitons of BiFeO$_3$, we introduce for each exciton $S$ an exciton-weighted density of states (WDOS) for the electron and the hole separately:
\begin{equation}
\begin{split}
    \mathrm{WDOS}_{S}^{\mathrm{elec}}(\omega) &= \frac{1}{N_{\mathrm{v}}}\sum_{\mathbf{k},c,v} \left| {A^{S}_{\mathbf{k},c,v}} \right|^{2}  \delta \left(\hbar\omega-\varepsilon^{(\mathrm{QP})}_{\mathbf{k},c}\right), \\
    \mathrm{WDOS}_{S}^{\mathrm{hole}}(\omega) &= \frac{1}{N_{\mathrm{c}}}\sum_{\mathbf{k},c,v} \left| {A^{S}_{\mathbf{k},c,v}} \right|^{2}  \delta \left(\hbar\omega-\varepsilon^{(\mathrm{QP})}_{\mathbf{k},v}\right),
    \label{eq:PDOS_exciton_weight}
\end{split}
\end{equation}
where $N_{\mathrm{c}}$ ($N_{\mathrm{v}}$) denotes the number of conduction (valence) bands used in the calculation of the excitons and we use the quasi-particle one-electron energies on the ev$GW$-level.
The two WDOSs are normalized such that the sum of the electron and hole WDOS, summed over all excitons, yields the canonical one-electron DOS: $\sum_S [\text{WDOS}^{\text{elec}}_S(\omega) + \text{WDOS}^{\text{hole}}_S(\omega)] = \text{DOS}(\omega)$.
The exciton-weighted DOS allows a simple, yet insightful visualization and classification of the constituent electron-hole pairs of an exciton in terms of their energy.
In Fig.~\ref{fig:t2g_eg_excitons}, we compare the WDOS for selected excitons to the ev$GW$-level DOS shown in Fig.~\ref{fig:t2g_eg_excitons}a.

As shown in our companion paper~\cite{aseem_sven_prl}, the exciton spectrum of BiFeO$_3$ consists of two distinct classes of excitons:
a set of 20 strongly bound excitons with binding energies in excess of 1.5~eV and a rest of only weakly bound excitonic states with binding energies of less than 0.3~eV.
We first focus on the strongly bound excitons, which all involve a flip of the electron spin, \textit{i.e.}, the expectation value of the spin operator in z-direction changes by approximately $\pm \hbar$.
These 20 excitons either come in exactly degenerate pairs that transform in the $E$-representation under the crystal space group, or come in \textit{de facto} degenerate pairs, the splitting of which is too small to resolve computationally and which transform in the $A_1$ and $A_2$ representation.
For simplicity, we will refer to these two different types of excitons as $E$-excitons and $A$-excitons, respectively.
We note that the $E$-excitons couple to light polarized in the x-y-plane, such as left- and right-circularly polarized light, while the $A$-excitons couple to light polarized along the z-axis.

In Fig.~\ref{fig:t2g_eg_excitons}b, we depict the WDOS of the 20 strongly bound excitons, 14 of which are $E$- and 6 of which are $A$-excitons.
We only show one member of each exact (or almost exact) doublet.
The WDOS plot reveals that, aside from their symmetry, the strongly bound excitons can also be distinguished by the contributing electron and hole states.
Twelve of the strongly bound excitons consist of transitions from the valence band top to the conduction band bottom (red lines in Fig.~\ref{fig:t2g_eg_excitons}b), while the other eight receive contributions from deeper within the conduction and/or valence band (blue lines in Fig.~\ref{fig:t2g_eg_excitons}b).
We can understand this partition as well as the $E$- or $A$-exciton character of these excitations within our symmetry-adapted orbital picture, factoring in SOC and selection rules.
We focus on our orbital picture of the PBE-level band structure first, as this gives already the right qualitative explanation of the strongly bound excitons and briefly comment on the quantitative changes induced by the ev$GW$ correction afterwards.

\begin{figure*}[]
   \centering
   \includegraphics[width=\textwidth]{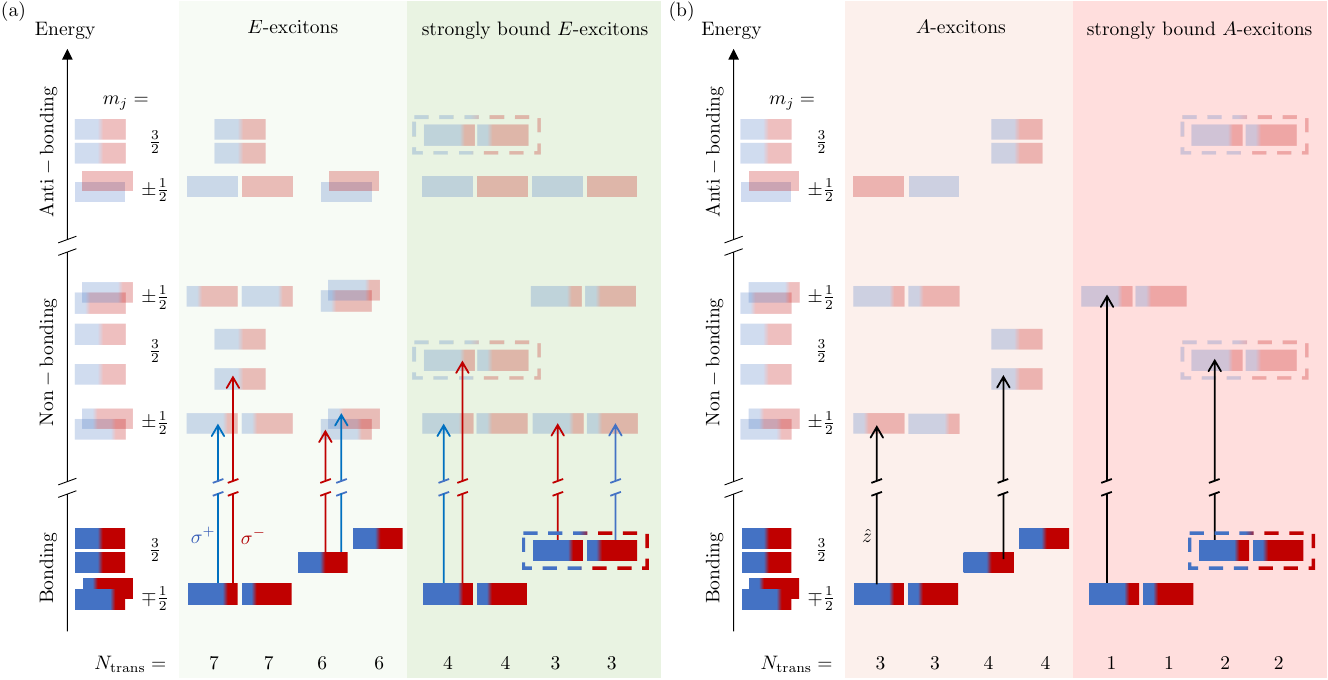}
   \caption{Electronic transitions between occupied and unoccupied Fe-$3d$ states in BiFeO$_3$, factoring in crystal field splitting and spin-orbit coupling, that make up the lowest-energy $E$- (panel a) and $A$-excitons (panel b).
   Left column: unoccupied (semi-transparent rectangles) and occupied (filled rectangles) Fe-$3d$ states with their total crystal angular momentum $m_j$.
   The colors of the rectangles schematically denote the weights of the contribution of each of the two Fe sites (see caption of Fig.~\ref{fig:crystal-field-splitting-scheme} for a more detailed description).
   Middle column: all possible electronic transitions from the four occupied bonding states to any of the non-bonding or anti-bonding states that are compatible with the angular momentum selection rules $\Delta m_j=\pm1$ for $E$- and $\Delta m_j=0$ for $A$-excitons.
   The arrows indicate examples of possible transitions, with blue (red) arrows corresponding to transitions that can be excited by $\sigma^+$- ($\sigma^-$)-polarized light and black arrows indicating transitions that are activated by light polarized along the $z$-axis.
   The numbers in the bottom row indicate the number of transitions compatible with the angular momentum selection rule.
   Right column: subsets and/or combinations of transitions in the middle column that give rise to a strongly bound exciton.
   Dashed boxes indicate the effective unmixing of the $m_j=3/2$-state pairs by virtue of the attractive electron-hole attraction overcoming the SOC-mediated state mixing.}
   \label{fig:transitions}
\end{figure*}

In Fig.~\ref{fig:transitions}, we show all possible transitions from the Fe-$3d$ states that form the top of the valence band to the Fe-$3d$ states that form the first conduction band and the bottom of the second conduction band.
The first column of Fig.~\ref{fig:transitions}a depicts the ten non-bonding and anti-bonding Fe-$3d$ states from Fig.~\ref{fig:crystal-field-splitting-scheme} as semi-transparent rectangles along with their total crystal angular momentum quantum number $m_j$ and the degree to which they are localized on either of the two Fe sites (blue and red shading, see also caption of Fig.~\ref{fig:crystal-field-splitting-scheme}).
In addition to the unoccupied states, we also include the four highest occupied Fe-$3d$ states, which are the counterparts of the unoccupied anti-bonding states and can be constructed in a similar fashion.
The degree of their localization on one or two Fe sites was determined from the orbital projections of the Kohn-Sham states of the PBE-level calculation under inclusion of SOC.
As in Fig.~\ref{fig:crystal-field-splitting-scheme}, rectangles stacked into the plane denote exactly degenerate states and the upper (lower) sign of the $m_j$ quantum numbers refers to the state in the front (back).

The second column of Fig.~\ref{fig:transitions}a depicts all possible transitions between the four different occupied states to any of the ten unoccupied states and that are compatible with the angular momentum selection rules for $E$-excitons, for which $\Delta m_j=\pm1$ for $\sigma^{\pm}$-circularly polarized light.
For illustration, we only explicitly illustrate four of the possible transitions by arrows, with a blue (red) arrow denoting transitions with angular momentum $\Delta m_j=+1$ ($\Delta m_j=-1$) that couple to $\sigma^+$- ($\sigma^-$-) light.
The $\Delta m_j = \pm 1$ selection rule allows a state with $m_j=-(+)1/2$ to couple to a state with $m_j=+(-)1/2$ and, notably, a state with $m_j=3/2$ to couple to \emph{both} states with $m_j=\pm1/2$ by virtue of the equivalence of $m_j=+3/2\equiv-3/2$ and vice versa.
Counting the possible transitions yields a total of 26 excitons that form 13 doublets of doubly degenerate excitons, as indicated at the bottom of Fig.~\ref{fig:transitions}a.

From these 26 excitons, we can now identify those that are strongly bound.
For this, we focus on the Fe site on which the occupied and the unoccupied states are localized.
A strongly bound exciton will arise if both the occupied and unoccupied state are localized on the same Fe site, as then the attraction between electron and hole is strongest, which results in a large binding energy.
For many transitions, this analysis is straightforward as both occupied and unoccupied state are already localized to a large degree on one Fe site only.
For those states that contain equal admixtures of the two Fe sites however, it is not immediately apparent if they can partake in the formation of a strongly bound exciton as they lack the localization of either the hole or the electron.
Notably, all of these maximally mixed states have the same quantum number $m_j=3/2$ and originate from the mixing of maximally \emph{localized} states by the spin-orbit interaction.
This suggests that states localized on predominantly one of the Fe sites can be formed by an effective \emph{unmixing} of these SOC-mixed states by the two-particle interaction that enters the Bethe-Salpeter equation.
These unmixed states can then form strongly bound excitons by binding to an electron or hole counterpart localized on the same Fe site.

We illustrate this idea in the third column of Fig.~\ref{fig:transitions}a, in which we depict the subset of all possible $E$-exciton transitions that result in a strongly bound exciton.
The mentioned effective unmixing of the $m_j=3/2$ states is indicated by a dashed box around each pair.
The strongly bound nature of these transitions arises as each hole state is paired with an electron state that is localized on the same Fe site.
In total, this method of analysis is able to account for all 14 strongly bound excitons of $E$-symmetry.
It moreover allows us the physical insight that, in order to give rise to a strongly bound exciton, the spin-orbit interaction has to be overcome by the electron-hole attraction, so that the SOC-mediated mixing of states can be undone.
More generally, this suggests that in materials with magnetic order, spin-polarized excitons can form only if the electron-hole attraction overcomes the spin-depolarizing spin-orbit interaction.

In a similar way, all strongly bound $A$-excitons can be explained and interpreted.
In Fig.~\ref{fig:transitions}b, we first construct for this in the second column all possible transitions from the four bonding states to the ten unoccupied Fe-$3d$ states that are compatible with the $\Delta m_j = 0$ selection rule of $A$-excitons, which couple to light polarized along the z-axis.
As a result of the $\Delta m_j = 0$ selection rule, the electron and hole of an $A$-exciton need to have the same angular momentum quantum number, \textit{i.e.}, only the transitions $m_j=\pm 1/2 \to \pm 1/2$ and $m_j=3/2 \to 3/2$ are allowed.
This selection rules results in 14 possible $A$-exciton transitions, two of which we explicitly marked with black arrows for illustrative purposes.

The angular momentum selection rule for $A$-excitons puts a severe constraint on the electron and the hole to form a strongly bound exciton, as many of the transitions shown in the second column of Fig.~\ref{fig:transitions}b feature an electron and hole that are localized on two \emph{different} Fe sites.
Factoring in the unmixing of the SOC-split $m_j=3/2$-states, we can use our symmetry analysis to identify three pairs of almost degenerate $A$-excitons, in excellent agreement with the results of our \textit{ab initio} calculation.
We note that, in contrast to the $E$-exciton states, which are exactly degenerate by virtue of the crystal symmetry, the $A$-excitons come in only \emph{almost} degenerate pairs, and should always be thought of as a pair of an $A_1$- and an $A_2$-exciton, using the representations of the point group.
However, for the strongly bound $A$-excitons, the splitting between the $A_1$- and $A_2$- members of each doublet is not discernible, as it arises from the coupling of the two localized excitations, which is much weaker than the on-site electron-hole attraction.
Lastly, we note that for two out of the three strongly bound $A$-exciton pairs, the SOC splitting of the electron or hole state has to be overcome by the electron-hole attraction.
This discussion thus shows that engineering localized electronic excitations of a desired symmetry necessitates a not too strong inter-site interaction and a not too strong proximity spin-orbit coupling from nearby atoms, as both of these scattering channels compete to its detriment with the attractive electron-hole potential. \par 

Qualitatively, the transitions of strongly bound excitons are usually thought of as single transitions from bonding states to non-/anti-bonding states, both of definite energy.
However, the excitonic WDOS plotted in Fig.~\ref{fig:t2g_eg_excitons}b, shows that quantitatively, states within an energy range of several eVs contribute to a single exciton.
The large smearing of the hole and electron WDOS contributing to the strongly bound excitons stems from their sharp spatial localization, which corresponds to a large spread of the exciton wave function in momentum space.
The momentum-space density of strongly bound excitons is plotted in Fig.~2a in our companion paper~\cite{aseem_sven_prl}, which shows a uniform distribution all over the first Brillouin zone.
By contrast, the WDOS of the weakly bound bright excitons (cf. Fig.~\ref{fig:t2g_eg_excitons}c) features sharp peaks in accordance with their localization in momentum space and their spreading in real space. \par 

In addition to the width of the energy band that contributes to a single exciton, the orbital character of the states contributing to an exciton is also qualitatively different when a more quantitatively accurate methodology is used.
Whereas in the semi-local DFT, there are two, separated conduction bands, the non-bonding and anti-bonding bands (\textit{c.f.} Fig.~\ref{fig:e_str_dos}b), the ev$GW$ corrections bring these bands much closer in energy (\textit{c.f.} Fig.~\ref{fig:e_str_dos}d) to a sizeable overlap of these bands on the energy scale.
While the majority of the strongly bound excitons shown in  Fig.~\ref{fig:t2g_eg_excitons}b can still be associated with electronic transitions to either n.-b. (red lines) or a.-b.(blue lines) states, this sizeable overlap between the n.-b. and a.-b. in terms of their energy also allow mixed excitons (marked by a purple asterisk), that defy the typical textbook crystal field-split excitations. \par 

This mixing of the n.-b. and a.-b. energy bands for particular excitons also leads to a counter-intuitive  ordering of the excitons on the energy scale.
Indeed, contrary to the ordering on the DFT-level, an a.-b. exciton at 1.25 eV  is the lowest-energy exciton state among the strongly bound excitons, which feature an exceptionally high binding energy of 3.34 eV.
This large binding energy is more than 1~eV more than the other strongly bound excitons (see Fig.~1f of Ref.~\cite{aseem_sven_prl}).
It is the direct result of the interference, \textit{i.e.}, a mixing between electronic excitations into the n.-b. and a.-b. states that are brought close to each other in energy by the many-body correlation energy, \textit{i.e.}, the ev$GW$ correction.
This lowest energy exciton has a natural partner in the form of the highest of the 20 strongly bound excitons at 2.69~eV, which also features a strong mixing between the n.-b. and a.-b. excitations.
Both the lowest and the highest strongly bound excitons are characterized by the electron-hole binding energy dominating the spread in independent electron-hole pair transitions.

Finally, we note that our predictions of exciton energies reconcile well with the results of past experimental studies.
For instance, several experimental studies have reported and identified weak optical features at 1.41~eV and 1.91~eV and hypothesized as strongly bound spin-flip excitons\cite{gomez-salces_effect_2012,meggle_temperature_2019,meggle_optical_pressure_2019,xu_optical_2009,Wei2017-bd,ramachandran_charge_2010}.
These features approximately coincide with the predicted set of excitons at $\sim$1.56 eV and $\sim$2.01 eV, respectively if typical temperature-dependent changes of electronic energies are factored in.
Our calculations also predict an exciton state at an energy below 1.25 eV (possibly around 0.82 eV when fully converged in the calculation).
Such a feature has indeed been seen in the optical absorption spectrum below 1~eV reported in Ref.~\cite{gomez-salces_effect_2012}.
Furthermore, signatures of the predicted excitons with energies $\sim$2.30~eV appear to show up as an enhancement of resonant Raman intensities~\cite{weber_temperature_2016,Cazayous2009-go}, in transient absorption measurements~\cite{Yamada2014-kk,Li2018-iv}, and in photoinduced absorption experiments~\cite{Burkert2016,meggle_temperature_2019,meggle_optical_pressure_2019}.
A more detailed comparison of our calculations with past experimental probes of optical spectra of BiFeO$_3$ can be found in Section~3 of the SM~\cite{si}.  \par 

\section{Summary and conclusions}

In this work, we have described the electronic structure of BiFeO$_3$ using \textit{ab initio} DFT and MBPT methods and have used them for a detailed analysis of the electronic and excitonic states.
We have combined our computational description with a symmetry analysis to arrive at an intuitive atomic-orbital picture of the electronic structure.
We find that the frontier electronic structure is dominated by Fe-$3d$ orbitals hybridized with O-$2p$ and Bi-$6p$ orbitals.
The splitting of the Fe-$3d$ orbitals in the crystal field of the oxygen atoms is arranged according to the local $C_3$ point group symmetry of one Fe site, giving rise to peculiar sets of six-lobed symmetry-adapted $d$-orbitals.
These sets of orbitals contribute to bonding and anti-bonding states when the lobes are pointing toward the O-atoms, and to non-bonding states when the lobes are pointing away from the O-atoms.
We explicitly showed that the mixing of hydrogenic orbital according to the $C_3$-symmetry yields one-electron orbitals that are in excellent qualitative agreement with those obtained \textit{ab initio}.
We further explained the DFT-level energy levels of these orbitals in detail by a qualitative analysis of the interplay of symmetry, the spin-orbit coupling, and the crystal field splitting.  \par 

Going beyond DFT, we employed the partially self-consistent ev$GW$ method to overcome the quantitative shortcomings of the semi-local DFT.
We find that the ev$GW$ method is sufficient and able to reproduce the experimentally measured band gap well and that it provides an altogether quantitatively accurate account of the electronic structure of BiFeO$_3$.
The renormalization of the density of states on the $evGW$-level leads to a more mixed character of the frontier conduction band edge with now substantial contributions from O-$2p$ and Bi-$6p$ states, in addition to the original Fe-$3d$ states.
We showed that that this renormalization is crucial for a correct description of the birefringence of the optical absorption spectrum and for the correct position of the absorption maxima as a function of light frequency. \par 

Beyond the independent-particle picture, we have used the BSE method to obtain bound excited states, \textit{i.e.}, excitons.
We find that the optical absorption spectrum of BiFeO$_3$ features strongly bound excitonic states deep within the primary band gap.
These excitons are localized on the Fe sites and can be thought of as the BiFeO$_3$ analogue of the $t_{2g} \rightarrow e_g$ $d$-shell transitions of octahedrally coordinated transition metals.
We characterize these strongly bound excitons in terms their of their symmetry character, their spin polarization, and the contribution they receive from hybridized electronic states in the crystal.
The $d^5$ antiferromagnetic order in BiFeO$_3$ further enforces a correlation between the Fe site-specific spatial localization of these excitations and their spin degree of freedom.
Moreover, we have shown that these excitons feature a complex composition and orbital energy ordering, contrary to their prevalent descriptions in terms of single electronic transitions between $3d$ states.
By computing a weighted density of states of the hole and electron components of these excitons, we have demonstrated that valance and conduction band states over a large energy window contribute to the formation of these excitons.
Lastly, we have found that these strongly bound excitons are spin-polarized and that this spin polarization is protected against the depolarizing spin-orbit coupling from neighboring Bi atoms by the large on-site electron-hole attraction.
In this regard, our qualitative analysis is able to not only account for all of these 20 strongly bound excitons, but also rationalize their symmetry characters and the angular momentum and spin contents. \par 
 
Finally, in our companion paper~\cite{aseem_sven_prl}, we have focused in more detail on the spin-flip excitons that possess $\pm \hbar$ angular momentum and that are localized on distinct Fe atoms associated with the two magnetic sublattices of BiFeO$_3$.
There we have shown that magnetic texture of the BiFeO$_3$ can be modulated by exciting these spin-flip excitons using circularly polarized light. \par 
In combination, we have thus presented a detailed \textit{ab initio} study of electronic and excitonic structure of BiFeO$_3$ and supplemented it with symmetry analysis, intuitive physical and chemical interpretations.

Going forward, we intend to explore the coupling of the explored electronic degrees of freedom with the lattice.
Such a coupling has been argued to be the source of observed large optical linewidths~\cite{xu_optical_2009,meggle_temperature_2019} and could also be responsible for the resonant enhancement of Raman scattering cross sections~\cite{weber_temperature_2016,Cazayous2009-go}.
We thus intend to explore this coupling in future work using the recently developed \textit{ab initio} method for resonant, one-phonon Raman scattering~\cite{Reichardt2019-bv,Reichardt2020-kc}. \par 
 
\FloatBarrier

\begin{acknowledgments}

We acknowledge funding from the National Research Fund (FNR) Luxembourg, project “RESRAMAN” (grant no. C20/MS/14802965).
The \textit{ab initio} calculation are performed on the University of Luxembourg's high-performance computing (ULHPC) clusters~\cite{Varrette2022-td}.
We thank Muralidhar Nalabothula and Ludger Wirtz for valuable input and fruitful discussions.

\end{acknowledgments}

\bibliographystyle{apsrev4-2}
\bibliography{refs}

\end{document}


\maketitle

\noindent\rule{\textwidth}{0.5pt}

\tableofcontents

\noindent\rule{\textwidth}{0.5pt}
\vspace{0.25cm}

\section{Convergence studies}

\begin{table*}[h]
\begin{center}

\begin{footnotesize}
\begin{tabular}{lll}
\hline \hline
\textbf{Parameter                                                                    } & \textbf{Value} & \textbf{{\sc Yambo} variable} \\ \hline \hline

PW energy cut-off PW for expanding the wavefunctions                                & 25 Ha   & \texttt{FFTGvecs}  \\
PW energy cut-off for calculation of dielectric screening ($GW$)                 & 6 Ha    &  \texttt{NGsBlkXp} \\
Number of bands used for calculation of correlation part of $\Sigma$          & 600    &  \texttt{GbndRnge} \\
Plasma frequency for Godby-Needs plasmon-pole approx.                         & 1 Ha    &  \texttt{PPAPntXp} \\
PW energy cut-off for calculation of dielectric screening (BSE)                  & 18 Ha    & \texttt{BSENGblk} \\
\hline \hline

\end{tabular}
\end{footnotesize}
\caption{Parameters used in BSE and $GW$ calculations.} \label{tab:yambo_parameter_4}
\end{center}

\end{table*}

 \begin{figure}[h!]
    \centering
    \includegraphics[width=0.55\textwidth]{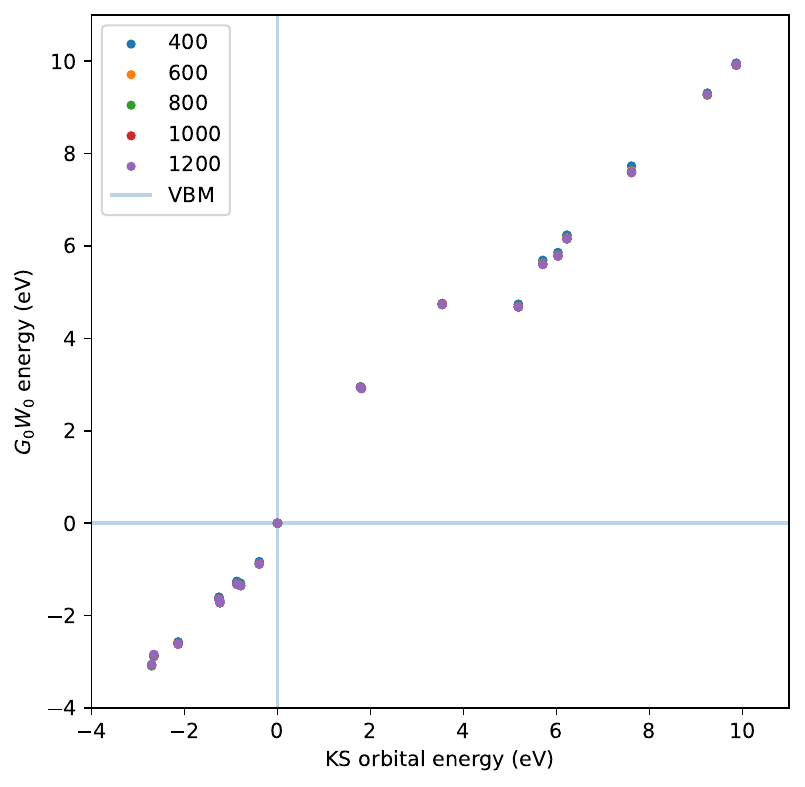}
    \caption{Convergence of $G_0W_0$ energies at $\Gamma$ with respect to number of bands considered in $G_0W_0$ calculation.}
    \label{fig:gw_bands_conv_si}
\end{figure}

 \begin{figure}[h!]
    \centering
    \includegraphics[width=0.55\textwidth]{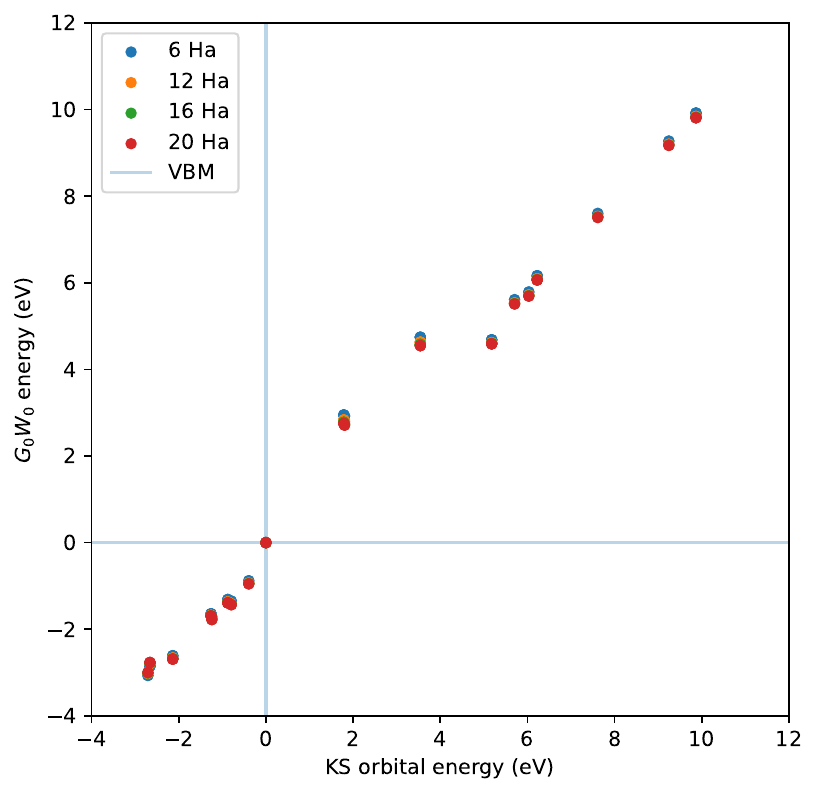}
    \caption{Convergence of $G_0W_0$ energies at $\Gamma$ with respect to plane-waves basis size used for calculation of dielectric response in $G_0W_0$.}
    \label{fig:gw_g_vec_conv_si}
\end{figure}

 \begin{figure}[h!]
    \centering
    \includegraphics[width=0.55\textwidth]{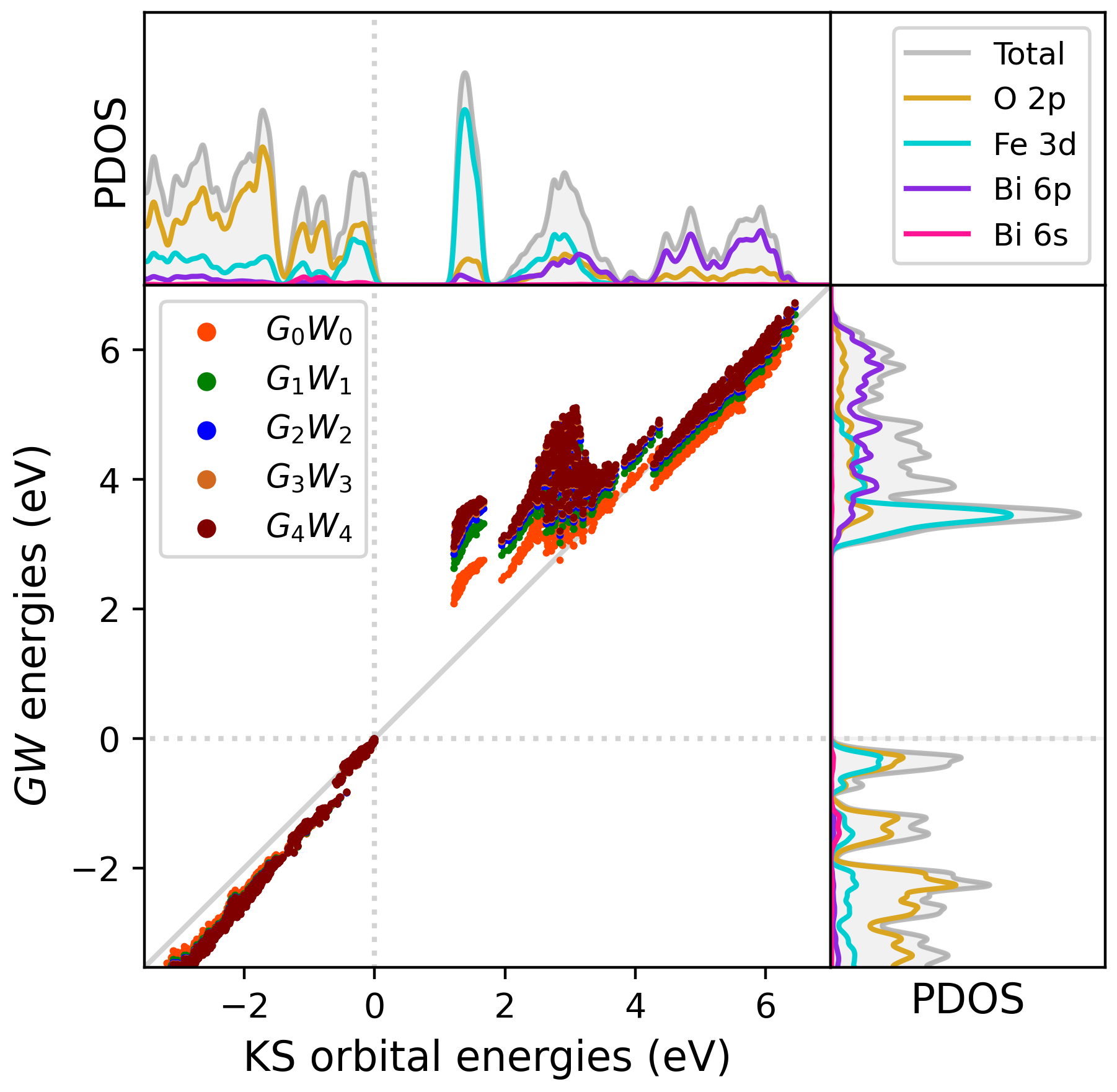}
    \caption{Convergence of $GW$ eigenvalues with respect to iterations of the eigenvalue self-consistent scheme.}
    \label{fig:gw_iter_conv_si}
\end{figure}

\begin{figure}[h!]
    \centering
    \includegraphics[width=0.55\textwidth]{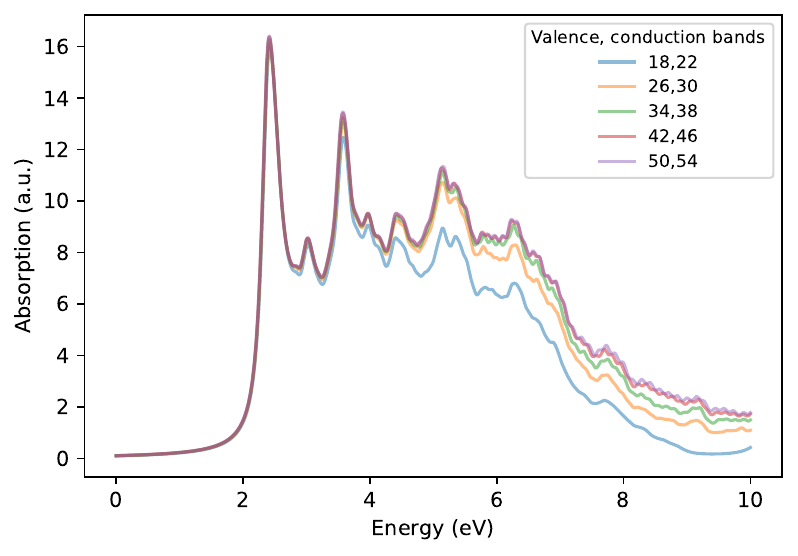}
    \caption{Convergence of BSE spectra with respect to size of transition basis. Calculations performed using a $4\times 4 \times 4$ k-mesh.}
    \label{fig:bse_bands_conv}
\end{figure}

\begin{figure}[h!]
    \centering
    \includegraphics[width=0.55\textwidth]{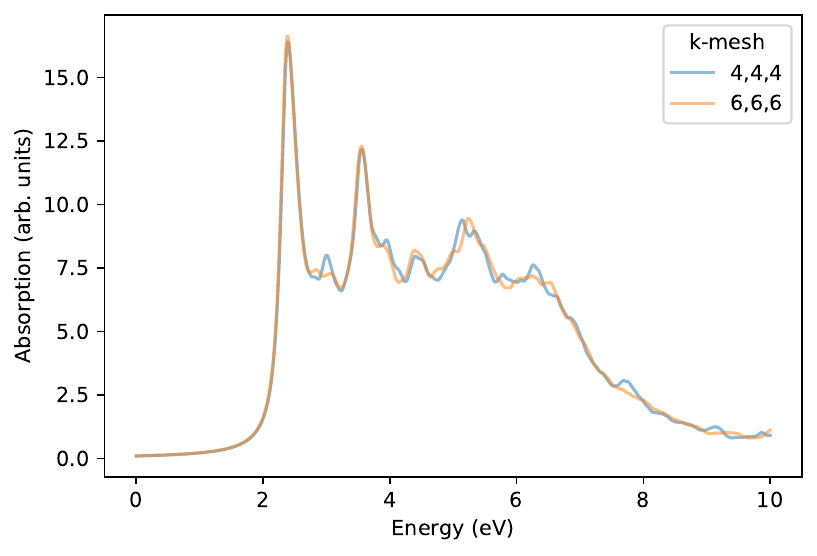}
    \caption{Convergence of BSE spectra with respect to k-point sampling.}
    \label{fig:bse_kmesh_conv}
\end{figure}

\begin{figure}[h!]
    \centering
    \includegraphics[width=0.55\textwidth]{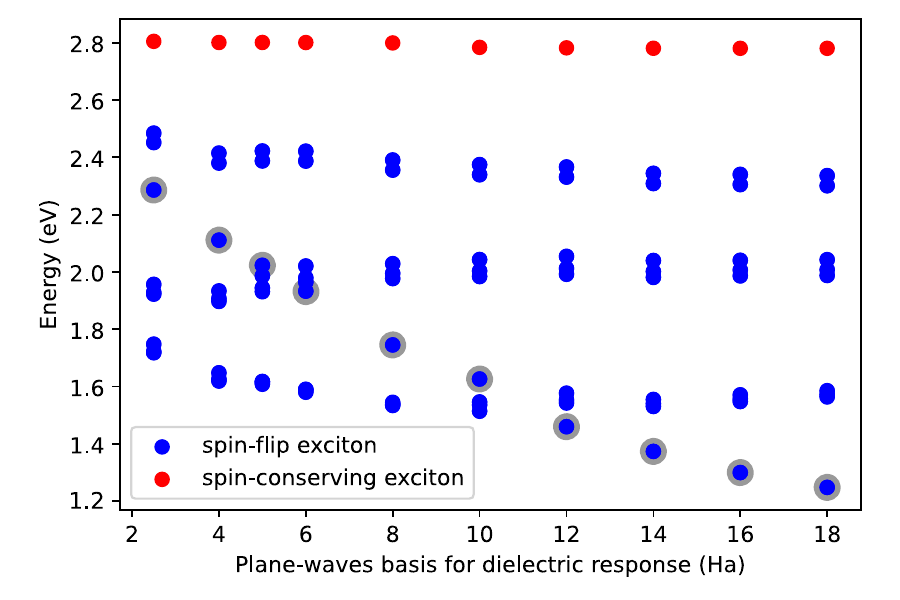}
    \caption{Convergence of BSE exciton energies with respect plane-wave basis for dielectric response. Exciton highlighted in gray circle is not fully converged with respect to the planewaves basis due to its exceptionally strong localisation.}
    \label{fig:bse_dielec_g_conv}
\end{figure}

\FloatBarrier
\newpage

\section{Additional figures}

 \begin{figure}[h!]
    \centering
    \includegraphics[width=0.5\textwidth]{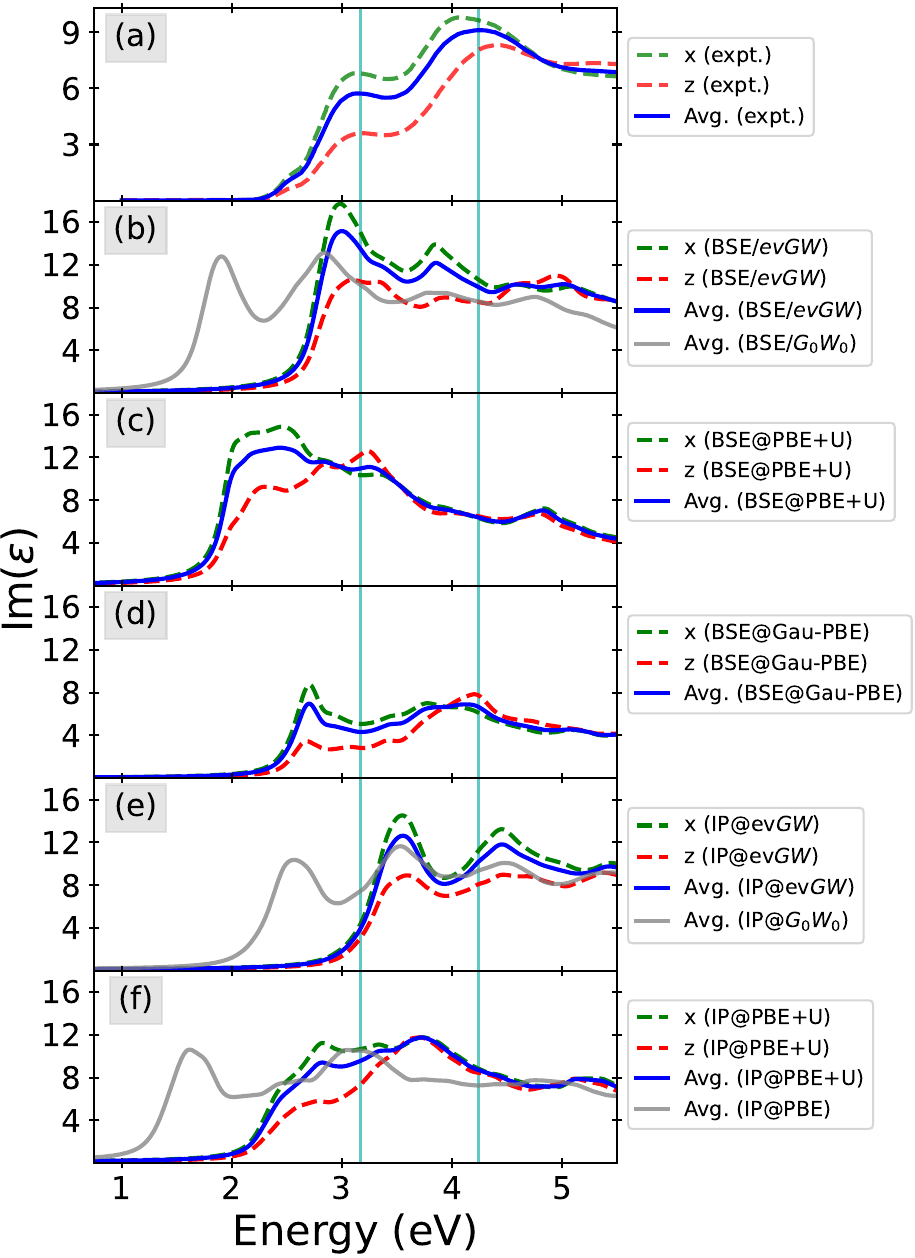}
    \caption{Optical spectra of BiFeO$_3$ calculated using different computational approaches. The experimentally measured spectra is taken from ref. \citenum{choi_optical_2011}. }
    \label{fig:optical_spectra_si}
\end{figure}

\begin{figure*}[ht!]
    \centering
    \includegraphics[width=0.98\textwidth]{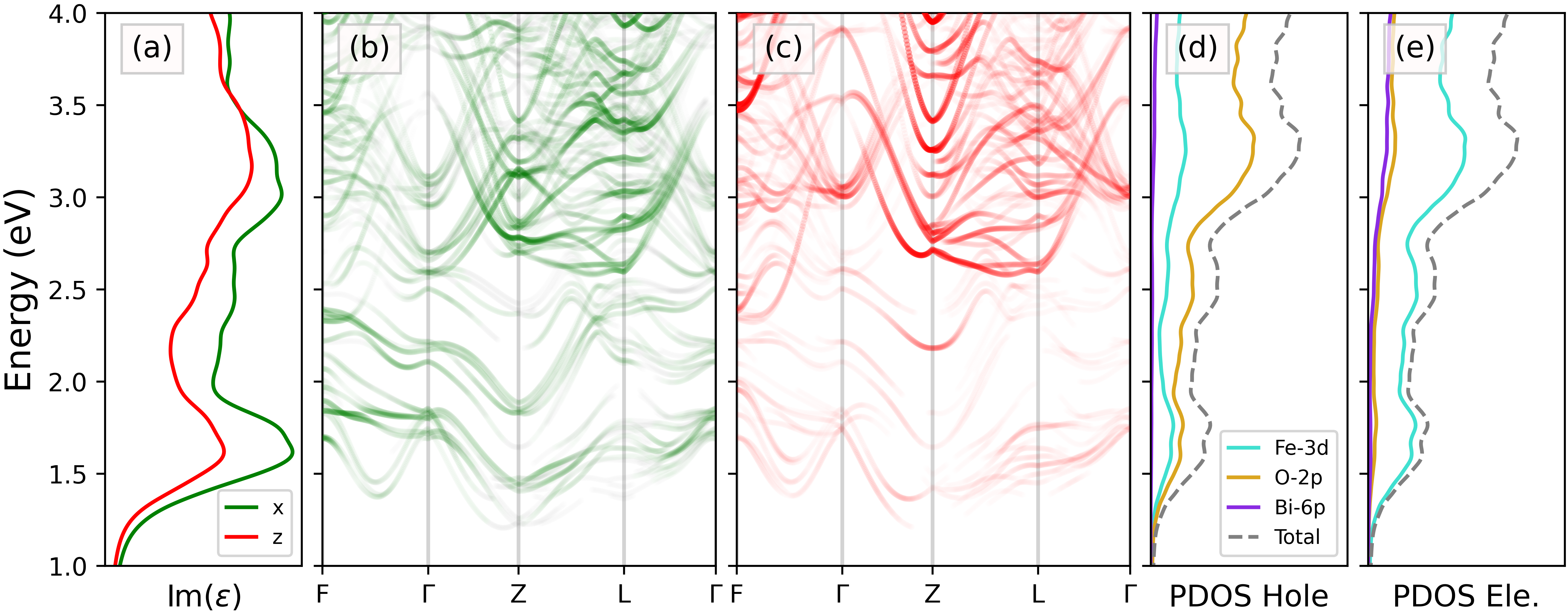}
    \caption{(a) Transition bandstructures (PBE) with opacity weighted by dipole matrix elements corresponding to light polarization along x-axis, (b) same as (a) but for light polarization along z-axis, (c) Im$(\epsilon)$, (d) and (e) density of transitions with corresponding hole and electron projected onto atomic orbitals.}
    \label{fig:optical_activity}
\end{figure*}

\FloatBarrier

\section{Comparison with past experimental probes of optical spectra of BiFeO$_3$}

Past experimental studies have reported and identified weak optical features at 1.41 eV and 1.91 eV as strongly bound spin-flip excitons\cite{gomez-salces_effect_2012,meggle_temperature_2019,meggle_optical_pressure_2019,xu_optical_2009,Wei2017-bd,ramachandran_charge_2010}. These features approximately coincide with the predicted set of excitons at $\sim$1.56 eV and $\sim$2.01 eV, respectively. Our calculations also predict an exciton state at energy \textit{below} 1.25 eV. Optical feature corresponding to this lowest energy exciton should be difficult to resolve experimentally considering much weaker optical absorption predicted by BSE (see fig.3b in \cite{aseem_sven_prl}). Features at $<$1 eV in the optical spectra reported in ref.~\cite{gomez-salces_effect_2012}  may correspond to the predicted lowest energy exciton state. \par 

Our calculations also predict in-gap spin-flip excitations at $\sim$2.30 eV.  There are several experimental signatures of an electronic state at this energy. Cathodoluminescence study on BiFeO$_3$ thin films by Hauser et al. found absorption features at 2.20 eV and 2.45 eV\cite{hauser_characterization_2008}. Hauser et al. ascribed the feature at 2.45 eV defect state, keeping the 2.20 eV feature open to interpretation. This feature may correspond exciton states at $\sim$2.30 eV as predicted by our calculations. Resonant enhancement of one-phonon Raman intensities have been reported for incident light of energy 2.18 eV and 2.33 eV~\cite{weber_temperature_2016}. Another study by Cazayous et. al reported two-phonon Raman resonance at 2.18 eV, 2.34 eV and 2.41 eV~\cite{Cazayous2009-go}. Both these studies confirm presence of electronic excited states near 2.30 eV, possibly corresponding to the predicted spin-flip exciton states.  \par

Experimental absorption spectra feature a shoulder at 2.45 eV to the peak at 3.0 eV\cite{choi_optical_2011,schmidt_anisotropic_2015,chen_optical_2010,xu_optical_2009,ramirez_magnon_2009,hauser_characterization_2008,basu_photoconductivity_2008,chen_optical_2010}. The presence of the shoulder has been associated in literature with double excitons, O-vacancy states\cite{radmilovic_combined_2020,clark_energy_2009,ju_first-principles_2009}, as well crystal field states. More recent state-of-the-art hybrid DFT calculations suggest that the O-vacancies contribute to the shoulder\cite{shimada_multiferroic_2016}. Apart from defect states, weakly bound exciton states in BiFeO$_3$ about 0.1-0.2 eV below the electronic gap, as predicted by our calculations may contribute to this shoulder at 2.45 eV. To accurately resolve the form of the optical spectra, especially at the absorption onset, it would be necessary to estimate the state-specific broadening of the exciton states, which is beyond the scope of this work.

Apart from direct absorption measurements, pump and probe methods can be used to discern and validate the in-gap exciton states predicted by our calculations.
In transient optical absorption measurements reported by Yamada et al.~\cite{Yamada2014-kk}, pumping electrons at 3.1 eV gives rise to change in optical density at $\sim$1.5 eV, $\sim$1.9 eV and $\sim$ 2.3 eV, possibly highlighting the predicted spin-flip exciton states. Signatures of excitonic states at 1.9 eV and 2.3 eV are also observed in a similar experiment on BiFeO$_3$ thin films by Li et al.~\cite{Li2018-iv}. Burkert et al. observed photoinduced features in the optical transmission spectra of BiFeO$_3$ at 1.22 eV, 1.66 eV, and 2.14 eV upon illumination with green laser (2.33 eV)~\cite{Burkert2016}. In a subsequent work on the temperature dependence of these optical features by Meggle et al., it is suggested that these features correspond to crystal field excitons consistent with their coupling with phonons\cite{meggle_temperature_2019}. This inference drawn by Meggle et al. is in alignment with our calculations. Further, Meggle et al. also showed that the positions of these optical features change linearly with compressive pressure, as characteriztic of onsite Fe$^{3+}$ excitations\cite{meggle_optical_pressure_2019}, matching with our computational characterization of the excitons . Overall, past experimental measurements are consistent with the our description of spin-flip exciton levels. \par

Our description of the exciton states in BiFeO$_3$ also shed light on the photostriction effect in BiFeO$_3$, e.g. light induced size change, as shown first by Kundys et al. in 2010~\cite{Kundys2010-fl}. This effect was explained by Kundys et al. as a combination of photovoltaic and inverse piezoelectric effect. The proposed mechanism was as follows: (1) light induced charge carriers separate due to intrinsic electric field, (2) the separated charges give rise to a photovoltaic potential difference, (3) this voltage gives rise to strain due to piezoelectric effect. Further experimental investigations of photostriction using time-resolved X-ray probes in combination with optical pump-probe methods revealed that response time associated with photostriction is of the order of nanoseconds~\cite{Li2015,Chen_2012,Lejman2014,Wen2013-eo}. It was also shown that photostriction effect involves a localized change in the lattice structure~\cite{Schick2014}. To explain the localized strain and the ultrafast response an excitonic mechanism of photostriction has been proposed~\cite{meggle_temperature_2019,Chen_2012}. This proposal is consistent with the fact that the photostriction effect in BiFeO$_3$ has also been observed for light irradiation of energy less than the band gap of BiFeO$_3$~\cite{PhysRevB.85.092301,Kundys2010-fl}. In the experiment of Kundys et al., the photostriction effect in BiFeO$_3$ induced using a 1.96 eV laser beam has been shown to decrease by up to 30$\%$ in presence of external magnetic field~\cite{Kundys2010-fl}. The decrease in the photostriction can be explained by change in the energies of spin-polarized excitations at $\sim$1.9-2.0 eV in presence of magnetic field. 

\begingroup\footnotesize
\let\section\subsubsection
\makeatletter
\renewcommand\@openbib@code{\itemsep\z@}
\makeatother
\bibliography{refs_sm}
\bibliographystyle{nature}
\endgroup